%% file: main.tex
\documentclass[fleqn,10pt]{wlscirep}
\usepackage[utf8]{inputenc}
\usepackage[T1]{fontenc}
\usepackage{graphicx}
\usepackage{subcaption}
\usepackage{algorithm}
\usepackage{algpseudocode}
\usepackage{tabularx}
\usepackage{subfiles}
\usepackage{amsmath, amssymb}
\usepackage{multirow}
\usepackage{hyperref}
\usepackage{soul}
\usepackage{booktabs} 

\usepackage[style=nature]{biblatex}
\usepackage{csquotes}
\title{Transabdominal Fetal Oximetry via Diffuse Optics: Principled Analysis and Demonstration in Pregnant Ovine Models}

\author[1, +, *]{Weitai Qian}
\author[1, +, *]{Rishad Raiyan Joarder}
\author[1]{Randall Fowler}
\author[1]{Begum Kasap}
\author[1]{Mahya Saffarpour}
\author[1]{Kourosh Vali}
\author[1]{Tailai Lihe}
\author[2,3]{Aijun Wang}
\author[3]{Diana Farmer}
\author[1]{Soheil Ghiasi}

\affil[+]{these authors contributed equally to this work}
\affil[1]{Department of Electrical and Computer Engineering, UC Davis, One Shields Avenue, Davis, CA, 95616, USA}
\affil[2]{Department of Surgery, UC Davis Health, 4301 X Street, Sacramento, CA, 95817, USA}
\affil[3]{Department of Biomedical Engineering, UC Davis, One Shields Avenue, Davis, CA, 95616, USA}

\affil[*]{\{wtqian, rrjoarder\}@ucdavis.edu}

\begin{abstract}
Diffuse optics has the potential to offer a substantial advancement in fetal health monitoring via enabling continuous measurement of fetal blood oxygen saturation (fSpO$_2$). Aiming to enhance the sensing accuracy and to elucidate the foundational limits of Transabdominal Fetal Oximetry (TFO) via diffuse optics, we introduce a theoretical derivation, and a comprehensive pipeline for fSpO$_2$ estimation from non-invasively sensed diffuse light intensity values, which are leveraged to analyze datasets obtained through both simulations and \emph{in-vivo} experiments in gold standard large animal model of pregnancy. We propose the Exponential Pulsation Ratio (EPR) as a key feature, and develop machine-learning models to fuse the information collected across multiple detectors. Our proposed method demonstrates a Mean Absolute Error (MAE) of 4.81\% and 6.85\% with a Pearson's r correlation of 0.81 (p<0.001) and 0.71 (p<0.001) for estimation of fSpO$_2$ in simulated dataset and \emph{in-vivo} dataset, respectively. Across both datasets, our method outperforms the existing approaches, enhancing the accuracy of the fSpO$_2$ estimation and demonstrates its viability as a supplemental technology for intrapartum fetal monitoring.
\end{abstract}

\begin{document}

\flushbottom
\maketitle
\thispagestyle{empty}

\section{Introduction}
\subfile{Sections/1_Introduction}

\section{Results}
\subfile{Sections/2_Results}

\section{Discussion}
\subfile{Sections/3_Discussion}

\section{Methods}\label{methods}
\subfile{Sections/4_Methods}

\section{Data Availability}
\subfile{Sections/5_Data_Availability}

\section{Code Availability}
\subfile{Sections/6_Code_Availability}

\printbibliography

\section{Acknowledgements}
\subfile{Sections/7_Acknowledgement}

\section{Author Information}
\subfile{Sections/8_Author_Information}

\section{Ethics Declarations}
\subfile{Sections/9_Ethics_Declarations}

\end{document}

%% file: Sections/1_Introduction.tex
\label{sec:introduction}

Cardiotocography (CTG), also known as Electrical Fetal Monitoring (EFM), is a noninvasive technology used to assess fetal well-being during labor and delivery. Developed about half a century ago, CTG relies on concurrent measurement of the fetal heart rate (FHR) and maternal uterine activity (UA) traces, which are subsequently interpreted by trained physicians to identify babies at risk of birth asphyxia, and complications associated with Hypoxic Ischemic Encephalopathy (HIE), such as cerebral palsy \cite{zhangBirthAsphyxiaAssociated2020, rainaldiPathophysiologyBirthAsphyxia2016}. Despite its intended benefits, the implementation of CTG has been associated with an increased rate of unnecessary interventions, such as emergency Cesarean section (C-sections) deliveries, and also potentially inappropriate or delayed interventions due to its high rate of false positives for detection of at-risk babies \cite{nelsonUncertainValueElectronic1996, Nelson_2016, Alfirevic_2017, BirthData_2020, Lancet_1985}. While standard of practice has been developed by leading professional organizations through extensive research \cite{ACOG_2009,FIGO_2015,NICE_2022}, CTG-based intrapartum fetal monitoring is not completely objective, as evidenced by the significant inter- and intra-observer variability in CTG trace interpretations \cite{Blix_2003, Nielsen_1987, Beaulieu_1982, Engelhart_2023}.

To address the low specificity associated with CTG for detection of babies at risk of HIE, a range of biochemical methods have been investigated, including blood pH sampling and lactate testing \cite{Young_2018, East2015, Irvine2009}. However, these methods are invasive, and do not offer continuous monitoring throughout labor and delivery. In response to the need for continuous and non-invasive assessment of fetal well-being, Transabdominal Fetal Oximetry (TFO) was developed as a tool to supplement CTG. This innovative system leverages the underlying principle of conventional pulse oximetry to non-invasively measure the arterial blood oxygen saturation of a fetus in utero, through the maternal tissue.

The TFO device emits two wavelengths of near infrared light to illuminate the  abdomen of a pregnant woman. The intensity of propagated light in tissue is modulated by the various tissue chromophores, with the degree of modulation depending on the light spectra, among other factors. The modulated intensity of diffuse light, often called a Photoplethysmography (PPG) signal, is captured by detectors positioned on the same side of the body as the light sources. This arrangement is referred to as reflectance pulse oximetry. TFO systems, which form the focus of this paper, are based on the principals of Continuous Wave Near Infrared Spectroscopy (CW-NIRS), and incorporate multiple detectors at separate Source-Detector Distances (SDDs)\cite{Fong_TECS, Fong_TBME}. The signals received from these systems are subsequently processed to continuously determine both the FHR and fSpO$_2$.

Alternative approaches to CW-NIRS, such as time-domain, frequency-domain or interferometric methods, exist, which yield more quantitative, photon time of flight distributions. However, such approaches support a far lower light throughput, and require expensive and highly sensitive measurement equipment, which may put them at a disadvantage from a clinical translation standpoint \cite{liuFiberbasedFrequencymodulatedContinuouswave2022, zavriyevRoleDiffuseCorrelation2021, zhaoQuantitativeRealtimePulse2018, hallacogluAbsoluteMeasurementCerebral2012, comelliVivoTimeresolvedReflectance2007a, quaresimaBilateralPrefrontalCortex2005, issOxiplexTSInfraredNoninvasive}. 

The processing pipeline to obtain fSpO$_2$ from measured PPG is based on the seminal work by Zourabian et al.\cite{zourabianTransabdominalMonitoringFetal2000}. This pipeline involves separating out the fetal AC component and a common DC component from the PPG signal, and taking the ratio of the fetal AC over DC for both wavelengths individually. This ratio is sometimes referred to as the Pulsation Ratio (PR). Next, one takes ratio of the two pulsation ratios at two wavelengths to obtain the Ratio-of-Ratios (RoR)\cite{severinghausTakuoAoyagiDiscovery2007}. Finally, RoR is mapped to fSpO$_2$ using either an equation\cite{zourabianTransabdominalMonitoringFetal2000} or a calibration curve.

\begin{figure}[tb!]
    \centering
    \includegraphics[width=\textwidth]{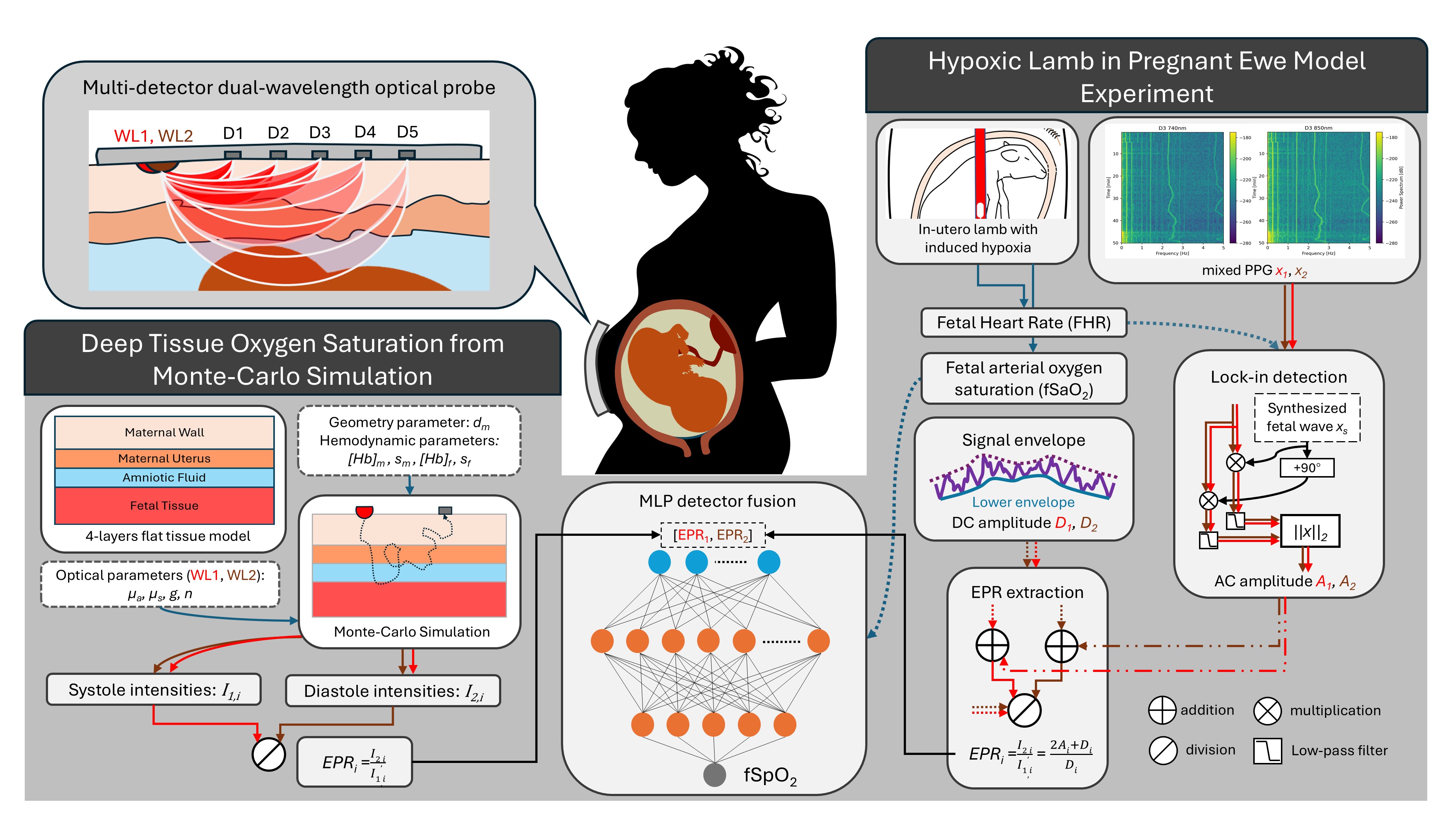}
    \caption{\textbf{A full schemes of deep tissue oxygen saturation sensing validated with dataset from simulation and experiment.} Exponential Pulsation Ratio from both dataset and validated with golden standard respectively. Details for each can be found under section \ref{sec:val_mc} and \ref{sec:val_sheep}}
    \label{fig:overview}
\end{figure}

Compared to conventional pulse oximetry, which is widely used and generally delivers accurate results, TFO faces additional complexities that are yet to be resolved. Firstly, the calibration of TFO systems presents significant challenges. Conventional transmittance pulse oximetry, where light sources and detectors are positioned on opposite sides of a peripheral body part with relatively thin tissue, relies on a few assumptions, such as uniform perfusion and uniform change to perfusion due to cardiac cycle, in the detected light field. Such assumptions considerably deviate from reality in reflectance-based measurement setup, which is used to non-invasively sense deep tissue in the TFO application. As a result, while transmittance pulse oximetry benefits from well-established and fairly accurate calibration methods, they do not readily apply to TFO, nor are they suitable for the lower oxygen saturation ranges of fetuses in utero \cite{Severinghaus_1990, Webb_1991}, where the average fSaO$_2$ levels were reported at or below 60\% \cite{Dildy_1994, Chua_1997}.

Moreover, TFO must isolate fetal signal from the measured mixed maternal-fetal PPG signals. Unlike conventional systems that collect signals from an individual, TFO must differentiate between contributions of the maternal and fetal tissue to the sensed signal, a challenging task given the weakness of the fetal signal, compared to the maternal signals and measurement noise floor. 

Finally, the depth of the pulsating artery significantly influences the parameters used to infer fSpO$_2$ from the measured PPG signals. While conventional pulse oximetry can disregard finger thickness variation as long as its assumptions are met, TFO cannot ignore the varying thicknesses of maternal abdominal tissues. Accurate calibration of TFO systems necessitates an estimation of the depth of fetal pulsating tissue, further complicating fSpO$_2$ inference process.

To overcome the outlined challenges and deliver a thorough theoretical and experimental analysis for fSpO$_2$ estimation using the TFO system (Figure \ref{fig:overview}), this paper contributes to the following objectives:
\begin{itemize}
    \item Elaborate on the existing theoretical framework for deep tissue reflectance oximetry via diffuse optics, with the objective of advancing the model, understanding and application.
    \item Validate and supplement our theoretical constructs with numerous Monte Carlo tissue simulations, and \emph{in-vivo} experimental data from large pregnant animal models.
    \item Investigate the specific challenges associated with TFO, and introduce a novel methodology for accurate fSpO$_2$ determination within this context.
\end{itemize}

%% file: Sections/2_Results.tex
\newcommand{\fcyL}{\mathcal{L}}
\newcommand{\EX}{\mathbb{E}}
\label{sec:results}

\subsection{Proposed Theoretical Framework}
In this section, we outline the theoretical foundation for estimating fSpO$_2$ and demonstrate that the Exponential Pulsation Ratio (EPR) is a superior feature compared to the conventional RoR model. Detailed derivations are provided in the \hyperref[methods]{Methods} Section.

The Beer-Lambert law in a highly scattering medium, like biological tissue, relates sensed light intensity to the absorption coefficient ($\mu_a$) of the medium. The normalized sensed intensity, or absorbance, is expressed as:
\begin{equation} 
    \text{Absorbance} = \frac{I}{I_0} = \int_{0}^{\inf} \exp(-\mu_a L) p(L) dL
    \label{eq:bll_base}
\end{equation}
where $p(L)$ is the pathlength probability density function (PDF) of light packets or photons within that medium, influenced by tissue geometry and scattering. Equation~\ref{eq:bll_base} can be extended to incorporate $M$ distinct mediums, where $f \in \{1, 2, \ldots, M\}$ represents fetal artery tissue:
\begin{align} \label{eq:bll_extd}
    \text{Absorbance} &= \int_{L_1} \int _{L_2} \ldots \int_{L_f}, \ldots, \int_{L_M} 
    \exp\left(-\sum_{j=1}^{M} \mu_{a,j} L_{j}\right) \notag p(L_1, L_2, \ldots, L_f, \ldots, L_M) \, dL_1 dL_2 \ldots dL_f, \ldots, dL_M \notag \\
    &= \int_{\fcyL} \exp\left(-\sum_{j=1}^{M} \mu_{a,j} L_{j}\right) p(\fcyL) \, d\fcyL
\end{align}
where $L_j$ and $\mu_{a,j}$ represent the optical pathlength and absorption coefficient of the $j^{th}$ tissue layer ($j \in \{1, 2, \ldots, M\}$), respectively. The PDF $p(L_1, L_2, \ldots, L_M)$ expresses the joint pathlength PDF. For compactness, we rewrite multiple integrals with $\fcyL$ and the joint PDF with $p(\fcyL)$. 

We assume that the measured intensity at the fetal peak (diastole) is $I_2$ and at the trough (systole) is $I_1$. From a frequency-domain perspective, these correspond to $DC + 2*AC_{fetal}$ and $DC$, respectively. The EPR per wavelength is defined as:
\begin{equation} \label{eq:epr1}
    \text{Exponential Pulsation Ratio (EPR)} = \frac{I_2}{I_1} = \frac{\EX_{L_f \sim p(L_f)}[\exp(-\mu_{a,f,2}L_f)]}{\EX_{L_f \sim p(L_f)}[\exp(-\mu_{a,f,1}L_f)]}
\end{equation}
where, $L_f$ is partial optical pathlength through pulsating fetal artery,  $\EX_{L_f \sim p(L_f)}$ denotes the expected value over the PDF $p(L_f)$, and $\mu_{a,f,1}$ and $\mu_{a,f,2}$ correspond to the fetal arterial absorption coefficients at $I_1$ and $I_2$, respectively. Note that both the PDF($p(L_f)$), and the $\mu_a$'s are functions of the source wavelength. 

From equation~\ref{eq:epr1}, the EPR term isolates fetal pulsating layer properties. With the fetal arterial $\mu_a$ isolated, albeit under certain conditions, EPR enables us to calculate fSpO$_2$ from a purer signal. The detailed derivation and comparison with the traditional RoR method can be found in section \ref{sec:theory}.

\subsection{Tissue Model Simulation}

\begin{figure}[tbh!]
    \centering
    \includegraphics[width=0.9\textwidth]{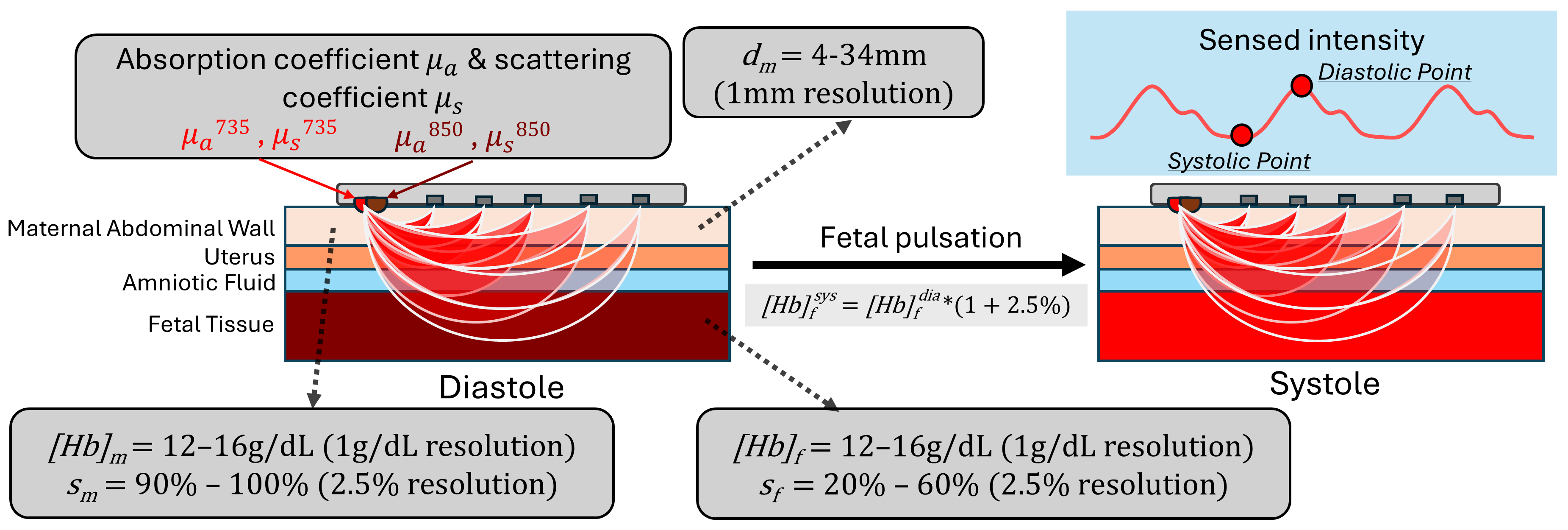}
    \caption{\textbf{The multi-layer tissue model used for MC simulation in this study.} It consists of one geometry parameter: Maternal abdominal wall thickness ($d_m$); four hemodynamic parameters: Maternal hemoglobin concentration ($[Hb]_m$), Maternal oxygen saturation ($s_m$), Fetal hemoglobin concentration ($[Hb]_f$), and Fetal oxygen saturation ($s_f$).}
    \label{fig:tissue_model}
\end{figure}

We employed Monte Carlo (MC) simulation, using a modified version of MCXtreme\cite{MonteCarloEXtreme2021}, to obtain static light intensity measurements from numerous tissue models. Our model comprises a flat, four-layered, dual-body structure in which, each layer utilizes homogeneous tissue properties. We deployed $10^{9}$ photons from two pencil beam light sources at wavelengths of 735nm and 850nm\cite{mannheimerWavelengthSelectionLowsaturation1997a}, activated sequentially and positioned directly above the model. The simulation incorporates 20 spherical detectors, each with a 2mm radius, arranged linearly from a 10mm to a 95mm Source-Detector Distance (SDD), with equal spacing between adjacent detectors. The tissue model consists of the following layers: 1.\textbf{Maternal Abdominal Wall} 2.\textbf{Uterus} 3.\textbf{Amniotic Fluid} 4.\textbf{Fetal Tissue}. Baseline dimensions and optical properties of the layers are based on Fong et al.\cite{fongContextuallyawareFetalSensing2020}. The geometry of the model is sketched in Figure ~\ref{fig:tissue_model}. 

To create a comprehensive simulation dataset, we evaluated a range of values for five parameters within our tissue model as illustrated in Figure ~\ref{fig:tissue_model}: 1.~\textbf{Maternal abdominal wall thickness} ($d_m$), 2.~\textbf{Maternal oxygen saturation} ($s_m$), 3.~\textbf{Maternal hemoglobin concentration} ($[Hb]_m$), 4.~\textbf{Fetal blood oxygen saturation} ($s_f$), and 5.~\textbf{Fetal hemoglobin concentration} ($[Hb]_f$). The first parameter modifies the model's geometry, while the subsequent parameters influence optical properties, specifically $\mu_a$, of their respective layers. Each wavelength employs a distinct set of optical properties, detailed further in Section~\ref{methods}.

To emulate fetal pulsation, two static tissue models associated with the peaks of fetal diastole and systole cycles are simulated. This approach suffices for calculating the EPR, thereby eliminating the need for simulation of intermediate tissue models. The two static models are obtained by varying the fetal hemoglobin concentration by 2.5\%\cite{reussMultilayerModelingReflectance2005} between the two states. That said, for each set of tissue parameters and wavelength, we conduct two static simulations. In total, over 2 million unique combinations of tissue model parameters, wavelengths, and systole/diastole states are simulated, providing a robust foundation for our analysis.

\begin{figure}[tbh!]
    \centering
    \includegraphics[width=\textwidth]{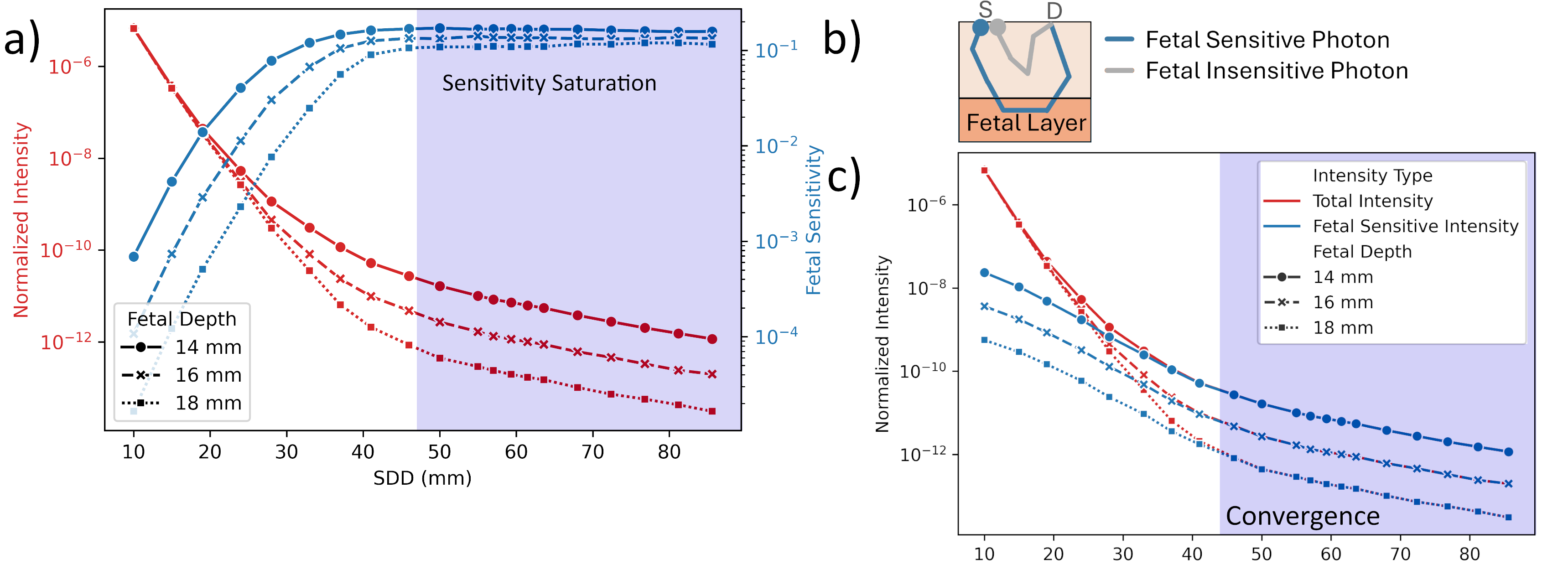}
    \caption{\textbf{Opposing changes in intensity and fetal sensitivity with SDD in simulated data.} (a) shows the normalized intensity decays exponentially with source detector distance (SDD) while fetal sensitivity increases to reach a saturation point. Simulation data generated with 735nm wavelength light source at three distinct fetal depths. (b) the classification of simulated photons into fetal sensitive/insensitive based on whether they probe the fetal layer. (c) shows the contribution of fetal sensitive photons to the overall intensity termed as Fetal Sensitive Intensity. Fetal Sensitive Intensity converges with Total Intensity beyond a certain SDD, showing that all photons become fetal sensitive. This explains the saturation of fetal sensitivity.}
\label{fig:sensitivity}
\end{figure}

\subsection{Normalized Intensity \& Fetal Sensitivity}
Normalized intensity ($\frac{I}{I_0}$) determines the best Signal-to-Noise Ratio (SNR) for a given SDD, while fetal sensitivity quantifies the portion of the signal from the fetal layer. Ideally, both fetal sensitivity and overall signal strength should be maximized; however, they exhibit opposing trends with SDD. This section explores these two criteria and their relationship with SDD.

First, we define \textit{Fetal Sensitivity} as the ratio between fetal to overall signal strength. Though the following equations take inspiration from recent simulation-based works on fetal saturation estimation, we shall apply our own theoretical framework to them \cite{wuSensitivityAnalysisTransabdominal2024b, guntherEffectPresenceAmniotic2021a}. Since PPG detects direct change in $\mu_a$, we quantify signal strength in terms of change in sensed intensity with respect to  change in $\mu_a$ ($\frac{\delta I}{\delta \mu_a}$). Since TFO is a body-in-body problem, the acquired PPG signal is a superposition both maternal and fetal signals. Assuming an ideal scenario where only the Maternal Abdominal Wall (Layer 1) pulsates, fetal sensitivity in our simulation setup is expressed as:
\begin{equation} \label{eq:fetal_sensitivity1}
    \text{Fetal Sensitivity} = \frac{\frac{\delta I}{\delta \mu_{a,f}}}{\frac{\delta I}{\delta \mu_{a,f}} + \frac{\delta I}{\delta \mu_{a,m}}}
\end{equation}
where $\mu_{a,f}$ and $\mu_{a,m}$ are the absorption coefficients of the Fetal and Maternal Abdominal Wall Layers, respectively. This equation can be further simplified via equation~\ref{eq:bll_extd} to:
\begin{equation} \label{eq:fetal_sensitivity2}
    \text{Fetal Sensitivity} = \frac{\int_\fcyL L_f \exp(-\sum_{j=1}^{M} \mu_{a,j}L_j)p(\fcyL)d\fcyL}{\int_\fcyL L_f \exp(-\sum_{j=1}^{M} \mu_{a,j}L_j)p(\fcyL)d\fcyL + \int_\fcyL L_m \exp(-\sum_{j=1}^{M} \mu_{a,j}L_j)p(\fcyL)d\fcyL}
\end{equation}
where, $L_f$ and $L_m$ are the optical pathlengths through the Fetal and Maternal Abdominal Layers, respectively.

We also defined \textit{Fetal Sensitivity Intensity}. In order to characterize this term, we classify the simulated photons into two groups: Fetal Sensitive and Fetal Insensitive. Photons that posses a non-zero optical pathlength within the fetal layer and ones that do not, as shown in Figure ~\ref{fig:sensitivity}b. Fetal Sensitive Intensity is the intensity contribution due to only fetal sensitive photons.

In Figure ~\ref{fig:sensitivity}, we plot normalized intensities versus SDD for a 735 nm wavelength at three fetal depths, alongside fetal sensitivity (a) and fetal sensitive intensity (c). While normalized intensity decays exponentially due to Beer-Lambert's Law, fetal sensitivity increases until reaching a saturation point. This saturation occurs because, at greater SDD, more photons penetrate the fetal layer. Beyond a certain SDD, all photons have already penetrated the fetal layer. This is further emphasized by the convergence in fetal sensitive and total intensity in part Figure ~\ref{fig:sensitivity}c. Whereas, in near detectors, fetal sensitive photons contribute very little to the overall signal. Also, note that the point of sensitivity saturation changes with fetal depth. 

Signal SNR and fetal sensitivity are opposing criteria that must be balanced; larger SDD improves fetal sensitivity but can reduce SNR. The optimal point is patient-dependent, offering key insights for TFO system design.

\subsection{Saturation-Geometry Ambiguity}

\begin{figure}[!tbh]
    \centering
    \includegraphics[width=\textwidth]{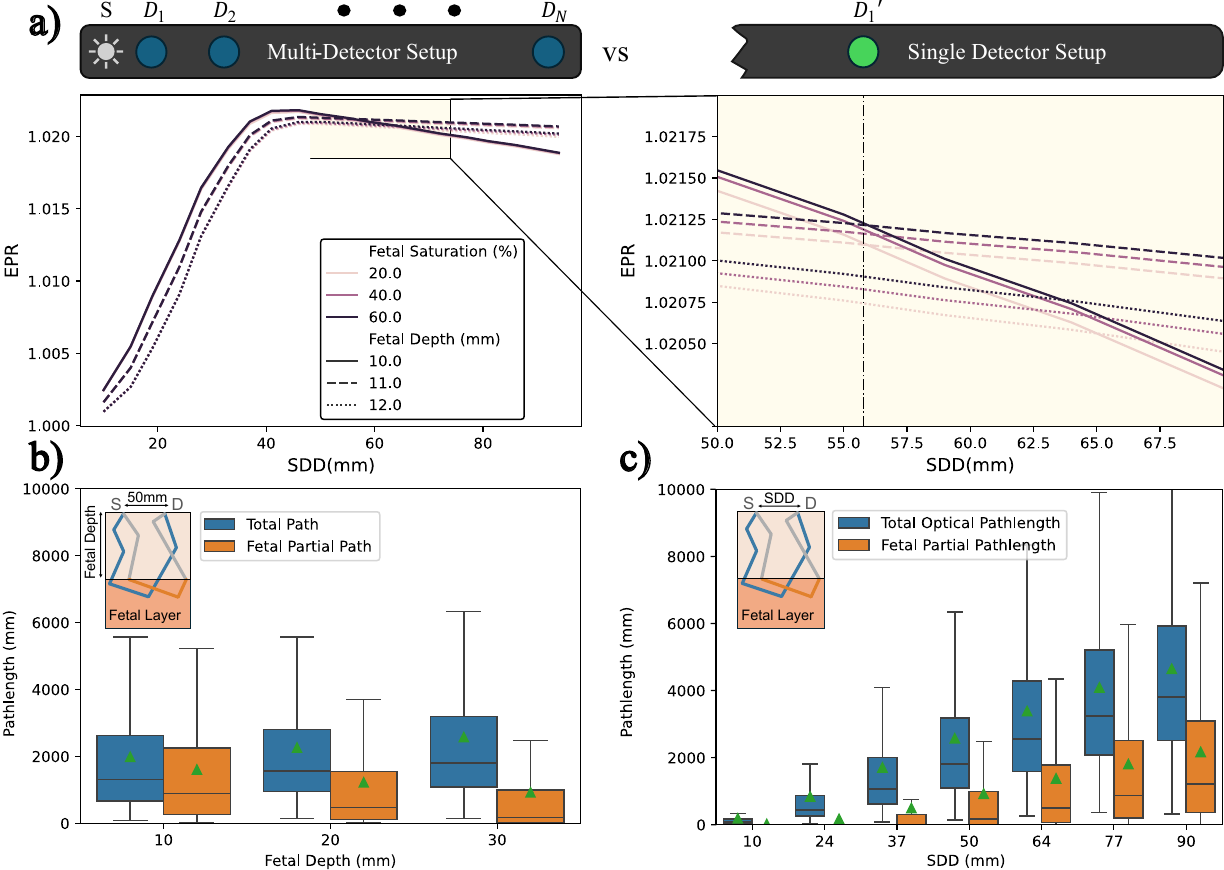}
    \caption{\textbf{A multi-detector setup can avoid saturation-geometry ambiguity in EPR, which occurs as a side-effect of sensed pathlength distribution.} (a) compares simulated single Wavelength EPR vs. SDD curves for 3 tissue geometries (fetal depth of 10, 11, \& 12mm) at 3 fSpO$_2$ levels (20\%, 40\%, \& 60\%). Left shows a multi-detector setup while right shows a zoomed in version of the curves in a single detector setup. A single detector along the vertical dash picks up the same EPR value for different saturation, fetal depth pairs, introducing ambiguity in EPR to fSpO$_2$ estimation. (b) \& (c) show photon's total and partial fetal pathlength distribution (mean, triangle) at different fetal depths (b) and SDD (c). The change in distribution due to depth causes the saturation-geometry ambiguity. Probing this distribution with a multi-detector setup promises a better approach.}
    \label{fig:saturation_geometry_ambiguity}
\end{figure}

From the perspective of a single photodetector, different fSpO$_2$ values can yield identical EPR values due to varying patient geometries, a phenomenon we term \textit{Saturation-Geometry Ambiguity}. This ambiguity affects conventional RoR measurements as well. Figure ~\ref{fig:saturation_geometry_ambiguity}(a) demonstrates this using simulation data from different embodiments of the aforementioned tissue model, where the bottom-most layer represents the fetus and the topmost layer, termed the Maternal abdominal wall, varies in thickness to represent different measurement geometries. For each geometry, different fetal oxygen saturation levels are simulated. Despite our use of a multi-detector setup, saturation-geometry ambiguity can be better understood in the context of a hypothetical single detector system. For example, a detector positioned at SDD of 55.5mm(vertical dashed line) records identical EPR values for different fSpO$_2$ levels at different measurement geometries. Thus, EPR derived from a single detector cannot provide a one-to-one mapping to fSpO$_2$ levels. 

However, the EPR vs. SDD curves distinctly vary across different \textit{saturation-geometry} pairs, suggesting that a more accurate fSpO$_2$ estimation is feasible via a multi-detector system. This observation underscores the advantage of employing multiple detectors, which can discern more nuanced fetal-induced patterns across different geometries, enhancing the reliability and accuracy of fSpO$_2$ estimations.

A theoretical grounding for a multi-detector setup can be established by looking at the simulated photon pathlength distributions. Figure ~\ref{fig:saturation_geometry_ambiguity}(b) shows the simulated photon's total pathlength distribution ($L_T$) and it's fractional component through the fetal layer ($L_f$) for different fetal depths at 50mm SDD. We show the four quartiles along with the mean (triangle). Returning to equation~\ref{eq:fetal_sensitivity2}, EPR is a function of the fetal pathlength PDF, $p(L_f)$. Which changes with fetal depth as shown in Figure ~\ref{fig:saturation_geometry_ambiguity}b. The change in $p(L_f)$'s PDF explains varying EPR values across different fetal depths for the same fetal saturation level. Generally speaking, any change in tissue geometry and/or scattering coefficient will affect this PDF. As a result, a single detector setup will measure different EPR for the same fSpO$_2$ across different patients. This is the root cause of the saturation-geometry ambiguity. However, this effect can be negligible if measurement geometry is similar across patients, as is the case for finger pulse oximetry. Deep-tissue systems overcome this by incorporating either spatial or temporal information to better capture the PDF, either in the form of spatially distributed detectors or Time-of-Flight (ToF) sensors. We show an example in Figure ~\ref{fig:saturation_geometry_ambiguity}c, which is similar to part Figure ~\ref{fig:saturation_geometry_ambiguity}b but with SDD along the x-axis. A Multi-detector setup collects more information about the fetal pathlength PDF, leading to a more accurate fSpO$_2$ estimation.

\subsection{fSpO$_2$ Estimation with Exponential Pulsation Ratios}
\subsubsection{fSpO$_2$ Estimation with Simulation Data}
\label{sec:result_simulation}
\begin{figure}[tbh!]
    \centering
    \begin{subfigure}[]{\textwidth}
        \includegraphics[width=\columnwidth]{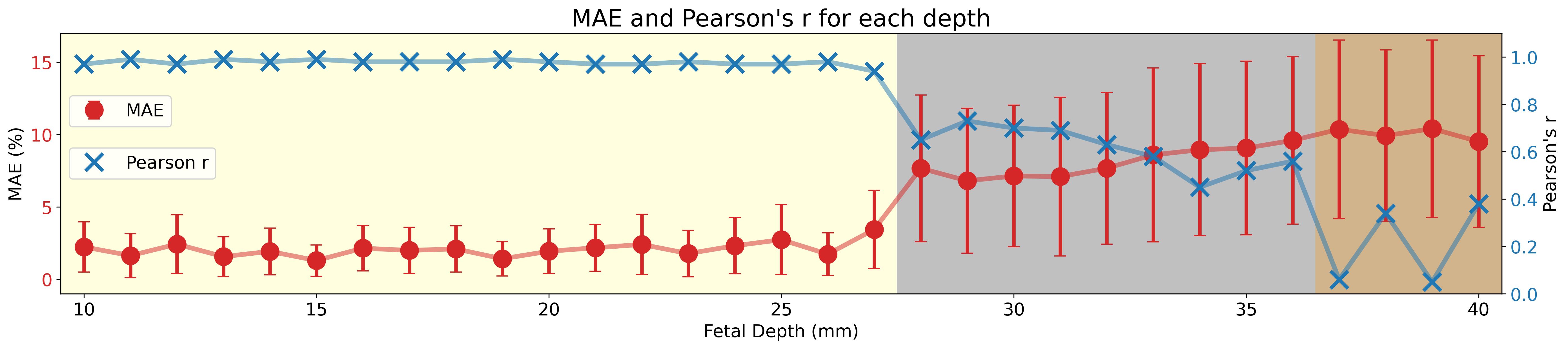}
        \caption{}
        \label{fig:mae_sim}
    \end{subfigure}
    \vspace{-1em}
    \begin{subfigure}[]{\textwidth}
        \includegraphics[width=\columnwidth]{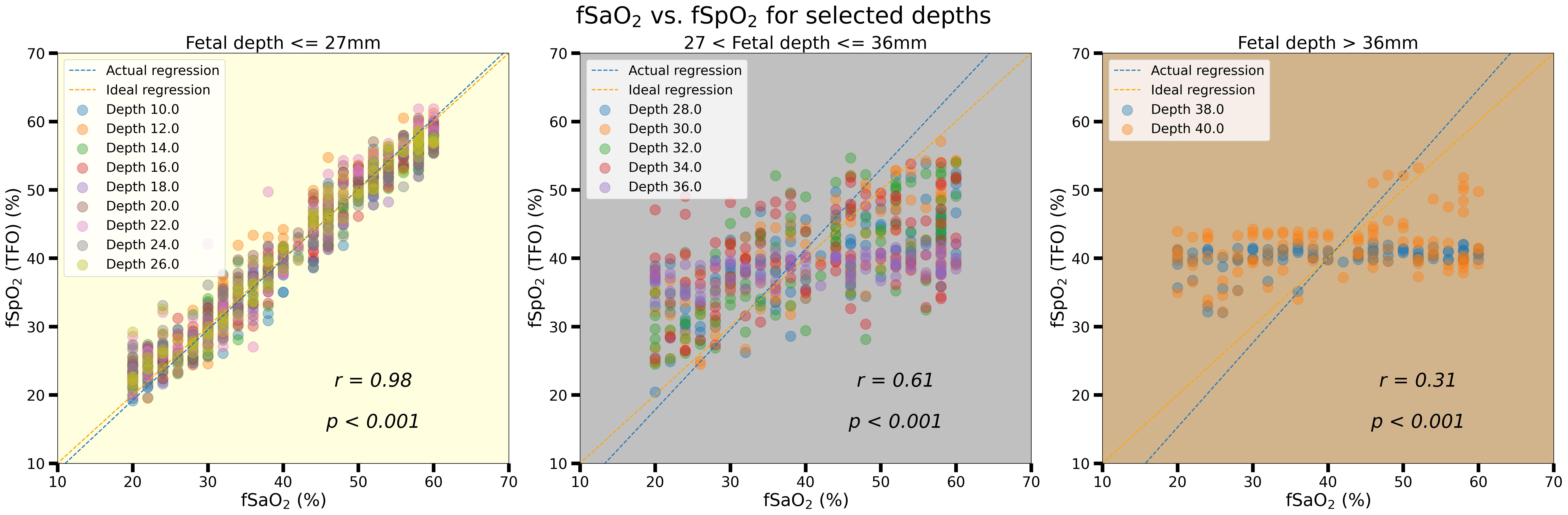}
        \caption{}
        \label{fig:corr_sim}
    \end{subfigure}
    \caption{(a) \textbf{Mean absolute error and Pearson's r correlation for fSpO$_2$ estimation at different depths.} All p-values for each depth were smaller than 0.001, so they are not annotated in the plot. Depths are partitioned into three groups based on the results. (b) \textbf{Pearson's r correlation between estimated fSpO$_2$ and actual fSaO$_2$ across selected simulation depths.} All depths are partitioned into three ranges and correlation were calculated separately. To enhance clarity and conciseness, results are presented only for even-numbered depths.}
\end{figure}

We validated our fSpO$_2$ estimation pipeline using data generated from MC simulations. Data from five SDDs at \{15, 33, 46, 68, 94\}mm were selected to strike a balance between representing a large range of SDDs and maintaining computational efficiency. Each sample comprised ten EPRs, calculated based on the light intensities received by five detectors at wavelengths of 735nm and 850nm (five EPRs per wavelength). A Multi-Layer Perceptron (MLP) network was utilized for fusion of information collected across the detectors. The dataset for each geometry was randomly split, allocating 80\% for training and 20\% for validation. Figure \ref{fig:mae_sim} displays the Mean Absolute Error (MAE) for the validation samples across all simulated depths, as well as Pearson's r correlation between the estimated fSpO$_2$ and the actual fSaO$_2$. The mean MAE across all depths was recorded at 4.84\%, with a trend of increase in errors and standard deviations for fetuses deeper than 26mm. It also reveals that Pearson's r exceeds 0.95 for shallow fetuses class (depth <= 27mm), while it decreases with the fetus depth increased.

Figure \ref{fig:corr_sim} illustrates the Pearson's r correlation when we partition all fetal depths into three ranges: 1) shallow fetus: depth <= 27mm; 2) moderate-depth fetus: 27 < depth <= 36mm; 3) deep fetus: depth > 36mm. The correlation is calculated with all validation samples in certain ranges of fetal depths, while the plot only visualize the samples from even depth for clarity. The correlation across all 31 depths is 0.81 with p < 0.001. For shallow fetuses and moderate-depth fetuses, it shows high or moderate correlation (r=0.98 and r=0.61) with p < 0.001. However, in the deep fetus class, the correlation is weak (r=0.31) with p < 0.001.

\subsubsection{fSpO$_2$ Estimation in Large Animal Models}


\begin{figure}[tbh!]
    \centering
    \includegraphics[width=\textwidth]{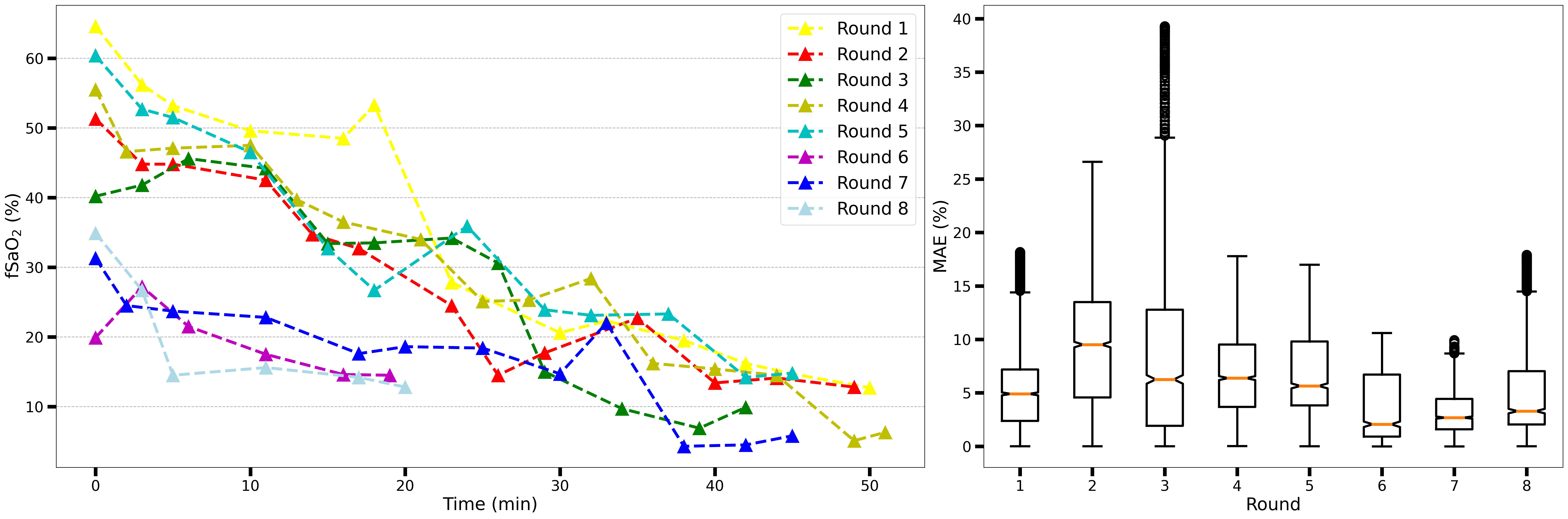}
    \caption{\textbf{Gold standard fSaO$_2$ readings from eight rounds of experiments and respective fSpO$_2$ estimation error.} Left: Temporal profile of fSaO$_2$ readings obtained via fetal blood gas analysis in eight hypoxic rounds of experiments utilized in this study. Right: Aggregated mean absolute error distribution of fSpO$_2$ estimation for eight hypoxic rounds in a 5-iteration cross validation.}
    \label{fig:mae_sheep}
\end{figure}

\begin{figure}[tbh!]
    \centering
    \includegraphics[width= \textwidth]{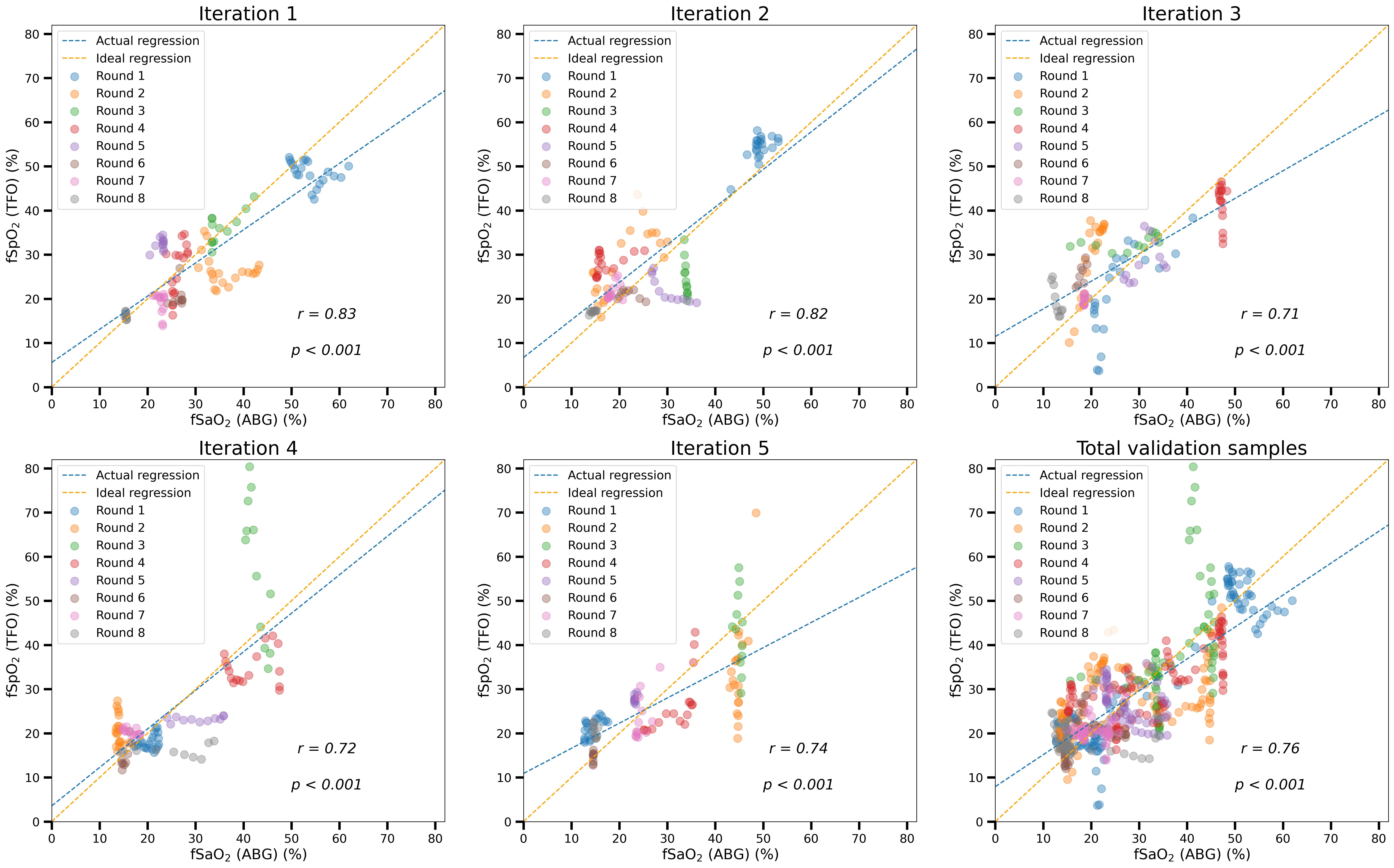}
    \caption{\textbf{Pearson's r correlation of fSpO$_2$ estimation for eight hypoxic rounds in a 5-iteration cross validation.} Validation samples from all eight rounds are displayed, with the first five subplots representing individual iterations of the cross-validation and the final subplot showing the aggregate results from all five iterations. For better visualization, only one sample for every 30 seconds is plotted, while the correlation is calculated for all validation samples.}
    \label{fig:correlation_sheep}
\end{figure}

We evaluated the effectiveness of employing EPRs for estimating fSpO$_2$ using data collected in gold standard large animal experiments (details in section~\ref{sec::lamb_exp}). The raw data is dual-spectrum (740nm and 850nm) PPG collected with five-detector optical probe each spacing at \{15, 30, 45, 70, 100\}mm to the LEDs. 

We utilized a MLP network for fusion of information across detectors. The network takes as input 10 EPRs calculated from the dual-wavelength mixed PPG signals. To mitigate bias and account for diverse sensing scenarios and geometries from different rounds, cross-validation was employed. In each iteration, data from each round was partitioned into training and validation samples following a temporal split scheme detailed in section~\ref{sec::lamb_fSpO2}. The validation data varied in every iteration for each round, ensuring comprehensive evaluation. The MAE for validation data from all rounds in each iteration is depicted in Figure ~\ref{fig:mae_sheep}. The weighted average MAE, adjusted by the number of samples in each round, across all iterations was [5.68\%, 6.47\%, 6.15\%, 6.60\%, 6.83\%], resulting in an average MAE of 6.35\% for all iterations and rounds combined. Notably, certain rounds exhibited consistently higher errors across all iterations, due to either lower signal quality or weak correlations between training and validation datasets.

Model performance was also evaluated using Pearson's correlation coefficient on validation data, as shown in Figure ~\ref{fig:correlation_sheep}. Each subplot compares estimated fSpO$_2$ with the gold standard fSaO$_2$ from blood gases, using 20\% of the data as validation and 80\% for training. All five iterations exhibit high correlations (r > 0.7, p < 0.001), with an overall correlation of r = 0.76 (p < 0.001) across all validation samples.

\subsection{Comparison between EPR and RoR}

\begin{table}[tbh!]
\centering
\begin{tabular}{c|cc|cc|cc}
\hline
\multirow{2}{*}{\textbf{Data Type}} & \multicolumn{2}{c|}{\textbf{Mean Absolute Error (\%)}}         & \multicolumn{2}{c|}{\textbf{Standard Deviation}}               & \multicolumn{2}{c}{\textbf{Pearson's r correlation}}           \\ \cline{2-7} 
                                    & \multicolumn{1}{c|}{\textbf{Simulation}} & \textbf{Experiment} & \multicolumn{1}{c|}{\textbf{Simulation}} & \textbf{Experiment} & \multicolumn{1}{c|}{\textbf{Simulation}} & \textbf{Experiment} \\ \hline
RoR                                 & 7.00                                     & 7.49                & 5.67                                     & 6.74                & 0.66                                     & 0.62                \\ \hline
EPR                                 & 4.81                                     & 6.85                & 5.14                                     & 5.94                & 0.81                                     & 0.71                \\ \hline
Improvement                         & 31.33\%                                  & 8.47\%              & 9.34\%                                   & 11.95\%             & 22.29\%                                  & 15.91\%             \\ \hline
\end{tabular}%
\caption{\textbf{Performance comparison for fSpO$_2$ estimation for validation data with RoRs and EPRs.}}
\label{table:ror_vs_epr}
\end{table}

To compare the performance of fSpO$_2$ estimation using the proposed EPR versus the conventional RoR features, identical preprocessing steps were applied. To comprehensively evaluate the effectiveness of EPR and RoR for estimating fSpO$_2$, we trained the same network, with different weight initialization, for five times and averaged the validation results across three metrics, summarized in Table \ref{table:ror_vs_epr}.

For the simulation data, the average MAE decreased from 7.00\% to 4.81\%, representing a 31.33\% improvement, while the standard deviation improved from 5.67\% to 5.14\%, and the correlation increased from 0.66 to 0.81, corresponding to improvements of 9.34\% and 22.29\%, respectively. In the animal model experiments, the MAE and standard deviation decreased from 7.49\% and 6.74\% to 6.85\% and 5.94\%, representing improvements of 8.47\% and 11.95\%. Similarly, Pearson's r correlation increased from 0.62 to 0.71, representing a 15.91\% improvement.

Averaging the results across five trials, these findings highlight the positive impact of utilizing EPR over conventional RoR for fSpO$_2$ estimation in both the simulation dataset, generated through Monte Carlo simulations, and the \emph{in vivo} dataset, collected from pregnant ewes with hypoxic lamb models.

\subsection{Performance Comparison on Unseen Noisy Simulation Data}
To better estimate the performance of our method in real-life scenarios, we re-evaluate our simulation results with added artificial noise. We specify two scenarios, only shot noise and shot noise paired with measurement noise from a gain amplifier. This setup is further explained in Section~\ref{sec:method_artificial_noise}. We compute MAE and error std. dev. using both our EPR approach and the conventional RoR. Our model is trained similar to Section~\ref{sec:result_simulation}.
\begin{table}[h!]
\centering
\label{tab:noise_data}
\begin{tabular}{ccccc}
\hline
\multirow{3}{*}{\textbf{Noise Type}} & \multicolumn{4}{c}{\textbf{Data Type}}                                \\ \cline{2-5} 
                                     & \multicolumn{2}{c}{\textbf{EPR}}  & \multicolumn{2}{c}{\textbf{RoR}}  \\ \cline{2-5} 
                                     & \textbf{MAE} & \textbf{Std. Dev.} & \textbf{MAE} & \textbf{Std. Dev.} \\ \hline
Shot Noise                           & 7.95\%       & 6.26\%             & 8.93\%       & 6.09\%             \\
Shot \& Measurement Noise            & 10.02\%      & 6.43\%             & 10.38\%      & 6.07\%             \\ \hline
\end{tabular}%
\caption{\textbf{Evaluation on noisy simulation data.}}
\end{table}

%% file: Sections/3_Discussion.tex
\label{sec:discussion}

This study highlights the limitation of straight forward extension of conventional pulse oximetry to  deep tissue reflectance settings, and proposes a principled approach to analysis and enhancement of sensing accuracy incorporating with a ML pipeline. We present comprehensive results including theoretical analysis, Monte Carlo simulation results, and experimental data from pregnant ewe with in-utero hypoxic lamb model.

We propose a novel per-wavelength Exponential Pulsation Ratio (EPR) derived from time-domain photoplethysmography (PPG) to estimate fetal oxygen saturation (fSpO$_2$). This approach offers a potentially superior alternative to the conventional Ratio-of-Ratios (RoR) method employed in both adult and fetal pulse oximetry \cite{chanPulseOximetryUnderstanding2013, zourabianTransabdominalMonitoringFetal2000, guntherReviewOpticalMethods2022b}. While RoR can be mathematically derived from the ratio of the logarithms of EPRs at distinct wavelengths, our work demonstrates the efficacy of a data-driven methodology for integrating these per-wavelength EPR features. This strategy affords machine learning models enhanced flexibility in mapping to fSaO$_2$ compared to the fixed structure of RoR, while concurrently incorporating sufficient physics-based constraints to mitigate overfitting. Integrating data from multiple detectors via a data-driven framework, in contrast to the conventional selection of a single optimal detector \cite{hielscherPhotonMigrationFetal2000, fongContextuallyawareFetalSensing2020}, our method enhances both accuracy and robustness in the estimation of fSpO$_2$ using per-wavelength EPR. Our approach includes an extensive Monte Carlo simulation of many tissue models, followed by extraction of EPRs from spatial distribution of simulated intensities. We also demonstrate the efficacy of the proposed approach for continuous fSpO$_2$ monitoring in a large pregnant animal model undergoing controlled fetal hypoxia.

Estimation of fSpO$_2$ from non-invasively measured intensity values using simulation data yields a MAE of 4.81\% across all fetal tissue depths, with errors increasing at greater depths. This degradation in accuracy is expected due to shorter photon path lengths in deeper fetal tissue, which limits the information in spatial intensities. At depths greater than 37 mm, the average MAE approaches or exceeds 10\%, resulting in poor correlation between estimated fSpO$_2$ and reference fSaO$_2$ values.

For the \emph{in-vivo} experimental data, we utilized a methodology similar to the one developed for simulation data and achieved a MAE of 6.85\%. Specifically, EPR are estimated from AC and DC components extracted through lock-in detection and lower signal envelope, respectively. Unlike the sensing noise-free settings used in Monte Carlo simulations, the task of fSpO$_2$ estimation from experimental data faces multiple practical challenges, leading to performance discrepancies as detailed in Table~\ref{table:ror_vs_epr} compared to fSpO$_2$ for simulation data. 

The quality of signals from experimental setups, typically quantified by SNR, is inherently lower than that from simulations. Simulations use static, homogeneous, flat tissue layers, while animal experiments involve dynamic, heterogeneous, and curved tissues. Factors such as probe alignment, maternal abdomen curvature, and the position and perfusion of the fetal tissue closest to the probe complicate data acquisition, leading to complex and noisy mixed PPG signals. It is not practical to reliably model or quantify the impact of all such variables throughout an experiment, contributing to complex and noisy acquired signal patterns compared to simulation data. Thus, this study trained and validated the model independently on simulation and in-vivo datasets due to their distributional differences. Future work will focus on bridging this domain gap. Pretraining the model using large-scale simulated data followed by fine-tuning with limited in-vivo data could enhance generalization. Approaches such as domain adaptation, semi-supervised learning, or unsupervised representation learning may further improve robustness across varying acquisition conditions.

Additionally, using a term-pregnant ewe model with hypoxic fetal lambs introduces further complexity for TFO. Experimental manipulations, such as surgical adjustments or pressure on the maternal abdomen, can alter measurement geometry and the optical properties of the tissue layers. Even minor pressure or manipulation may affect the acquired spatial intensity signals, potentially modifying tissue structure under the probe. Given the subtle influence of fSpO$_2$ on these signals, small perturbations can lead to significant errors in sensing accuracy.

\begin{figure}[tbh!]
    \centering
    \includegraphics[width=0.7\textwidth]{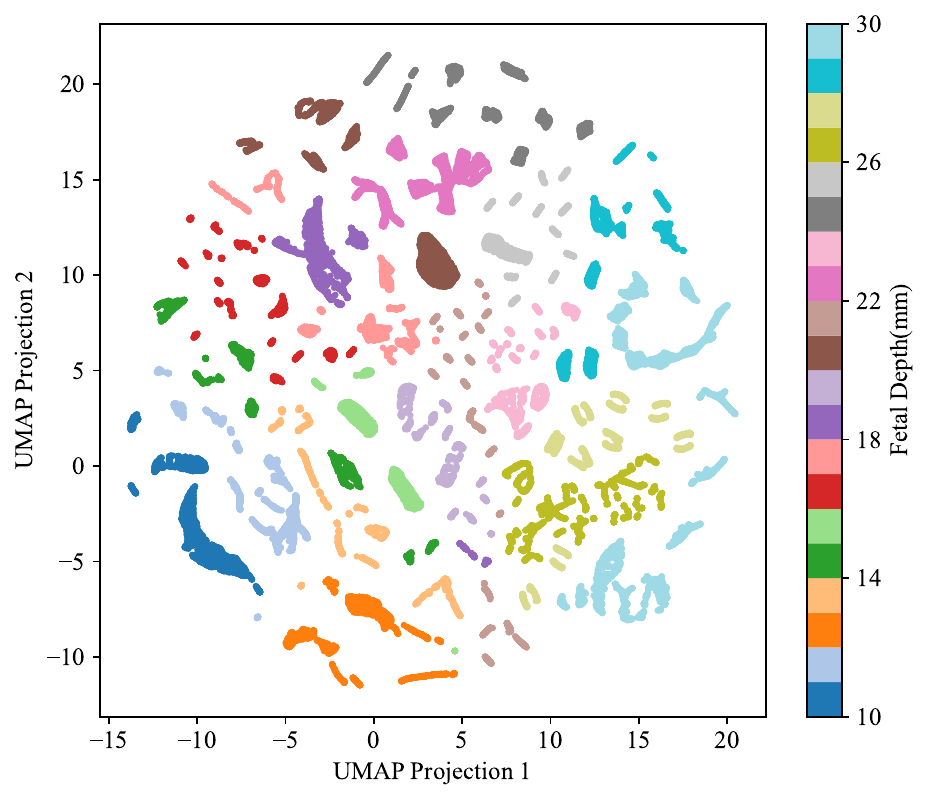}
    \caption{\textbf{UMAP projection of simulated 10-dimensional EPR feature vectors color-coded by fetal depth.} Each 10-dimensional EPR vector (5 detectors × 2 wavelengths) was projected onto a 2D plane using UMAP, preserving the high-dimensional clustering of data points, for easier visualization. Data points corresponding to distinct fetal depths appear as disjoint clusters, thus illustrating the challenge of generalizing to unseen fetal depths due to the extrapolation nature of the problem.}
    \label{fig:umap_projection_alt}
\end{figure}

Our models exhibited limited generalization to novel patient geometries, as observed in both the in-vivo sheep data across different subjects and the simulation dataset across varying fetal depths. In the simulation setting, each input was represented by a 10-dimensional EPR vector, formed by concatenating EPR values across five detectors and two wavelengths. These high-dimensional vectors were projected into a two-dimensional space using Uniform Manifold Approximation and Projection (UMAP) \cite{mcinnes2018umap}, a non-linear dimensionality reduction technique that preserves local data structure. The resulting 2D embedding, shown in Figure~\ref{fig:umap_projection_alt}, reveals distinct clusters corresponding to different fetal depths, generated from 20 uniformly spaced depths with 5-fold subsampling (UMAP parameters: \texttt{n\_neighbors=5}, Euclidean distance). This clustering pattern suggests that the EPR vectors are highly sensitive to geometric variation. Thus, predicting fSpO$_2$ for previously unseen fetal depths constitutes an extrapolation problem, a task known to be challenging for most statistical learning algorithms. Similar limitations in generalization across geometries have been reported in other studies \cite{guntherReviewOpticalMethods2022b}.

In this study, to explore the feasibility of deeper fetal sensing in human-relevant conditions, we modeled a wide range of the top-layer thickness of the four-layer Monte Carlo tissue model, effectively simulating thicker maternal abdominal walls compared to the ewe model. This allowed us to investigate the sensing limits in idealized human-like anatomical scenarios.

In clinical practice, acquiring large volumes of labeled fetal oxygenation data is not feasible. However, small numbers of reference samples--such as umbilical cord blood gas at delivery or fetal scalp blood sampling during labor--could serve as calibration points. These sparse but reliable measurements may enable personalized estimation through one-shot or few-shot learning, reducing the need for extensive labeled data while improving model adaptability.

These factors highlight the challenges and future directions in translating TFO technology from bench to bedside. Realizing the clinical potential of TFO will require future efforts to enhance simulation realism, develop data-efficient learning frameworks, and integrate sparse but reliable clinical calibration strategies. These advances will be essential for eventually enabling robust, generalizable, and non-invasive fSpO$_2$ monitoring in human subjects.

%% file: Sections/4_Methods.tex
\newcommand{\fcyL}{\mathcal{L}}
\newcommand{\EX}{\mathbb{E}}

\subsection{Theoretical Framework}\label{sec:theory}
\subsubsection{Conventional Approach to Fetal Pulse Oximetry}
The conventional method for fetal pulse oximetry is based on the seminal paper by Zourabian et al.\cite{zourabianTransabdominalMonitoringFetal2000} published in 2000. The paper takes advantage of the Modified Beer-Lambert law to establish a relationship between measured intensity and fSpO$_2$. This equation hinged on a term called Ratio of Ratio (RoR), which refers to the ratio between fetal AC components at two different wavelengths from the PPG signals, with each AC term normalized by its corresponding DC component. In case of single-body pulse oximetry, a straightforward derivation relates the RoR to fSpO$_2$.

There was however one parameter that needed to be calibrated for this equation. Dubbed as the pathlength factor, the wavelength-dependent parameter depends on the patient tissue composition and detector placement. Pathlength factor serves to explain the penetration depth difference between the two wavelengths and can vary significantly between different patients. There have been many attempts to calibrate for the pathlength factor. These include utilizing the wavelength dependency of the pathlength factor \cite{wuSelfcalibratedPulseOximetry2023}, using wavelengths with similar pathlength distributions\cite{nitzanCalibrationFreePulseOximetry2014}, using linear regression models\cite{valiEstimationFetalBlood2021}, among other attempts.

\subsubsection{Limitation of the Conventional Approach}
The shortcoming of the conventional approach stems from treating the detected light energy as having traveled over an average optical pathlength, rather than a collection of photons with different pathlengths. This assumption is particularly problematic in highly scattering media such as human tissue, especially when considering large optical pathlengths, as is the case with TFO.

We delved into understanding the implication of this assumption by first examining the modified Beer-Lambert law, which relates the ratio of sensed intensity to the input intensity with the absorption coefficient of the medium:
\begin{equation} \label{eq:section2eq0_1}
    \frac{I}{I_0} = \exp(-\mu_a \hat{L} + G)
\end{equation}
where, $I$ is the sensed intensity, $I_0$ is the radiated intensity into the medium, $\mu_a$ is the absorption coefficient and $\hat{L}$ is the mean optical pathlength. $G$ is a term that accounts for scattering loss, and other geometric factors. 

This equation is derived based on implicit assumptions that break in the case of TFO, mainly due to the highly scattering nature of human tissue coupled with the large depth of the tissue layer of interest, i.e., the fetus. These conditions give rise to photons with large optical pathlengths that undergo multiple scattering events before hitting a photodetector.

\subsubsection{Wide Distribution of Photon Pathlength}
The highly scattering nature of tissue in near infrared wavelengths coupled with large source-detector distance result in a wide distribution for the pathlength of detected photons. Consequently, the distribution cannot be adequately modeled with its mean. That is, a good representation would need to sufficiently incorporate the heterogeneous nature of detected photon-tissue interactions. We propose to extend the Beer-Lambert law via the following integral form to address the diverse mix of detected photon pathlengths:
\begin{equation} \label{eq:section2eq0_2}
    \frac{I}{I_0} = \int_{0}^{\inf} \exp(-\mu_a L) p(L) dL
\end{equation}
Here, $L$ is the optical pathlength for each detected photon, $p(L)$ is its density function. $p(L)$ is a function of the patient geometry and scattering properties of the tissue. Note that in a strict sense, the above equation should have been a summation due to the quantum nature of light. Assuming the number of detected photons are large, an integral form is used to approximate the summation with negligible loss of accuracy. This assumption is typically valid for Continuous Wave Near Infrared Spectroscopy (CW-NIRS) systems.

While equation~\ref{eq:section2eq0_1} serves as a simplification of~\ref{eq:section2eq0_2}, the probability density function $p(L)$ can vary significantly across different patients due to varying geometry and tissue scattering properties, factors that are especially pronounced in TFO due to the deep positioning of the fetus and the surrounding amniotic fluid. Therefore, the simplified exponential model does not adequately capture the complex interactions present in TFO.

Further down the derivations, the convention is to take the logarithm of $\frac{I}{I_0}$. In the simplified case, this would cancel the exponent, and result in a linear right-hand side. However, in the multi-photon case, this logarithm is applied to an integral of exponential functions, which introduces additional complexity to the model.

A similar analysis can demonstrate that the fetal RoR does not simplify calculations but adds complexity, further validating the need for this refined approach in TFO estimations.

\subsubsection{Theoretical Framework for Transabdominal Fetal Oximetry}
We initiate our theoretical framework by extending equation~\ref{eq:section2eq0_2} for a multi-tissue medium case, assuming the patient's body comprises $M$ homogeneous tissue layers. We apply a similar treatment as Sassaroli et al. to the Beer-Lambert Law and replace time-of-flight with optical pathlength and extend to multi-layers \cite{sassaroliCommentModifiedBeer2004}. But unlike this approach, we never introduce the notion of expected value of pathlenth($<L>$) for any of the layers. Rather, we leave the entire exponential term intact. We believe that a cause of inaccuracies with the current derivations is assuming a Gaussian distribution on the partial pathlengths to simplify the derivations. However, our Monte Carlo simulations often show non-Gaussian, distinct, geometry-dependent distributions. This is especially true for deeper tissue models under reflectance pulse oximetry. Thus, our derivation we attempt a more generalized solution which does not assume a known distribution. With all that said, the sensed intensity at a specific detector can be written as follows:
\begin{equation} \label{eq:section2eq1}
    \text{Absorbance} = \frac{I}{I_0} = \int_{L_1} \int _{L_2} \ldots  \int_{L_f}, \ldots, \int_{L_M} \exp(-\sum_{j=1}^{M} \mu_{a,j} L_{j}) p(L_1, L_2, \ldots,  L_f, \ldots, L_M ) dL_1 dL_2 \ldots  dL_f, \ldots, dL_M
\end{equation}
where, $\mu_{a,j}$ is the wavelength-dependent absorption coefficient of the $j^{th}$ tissue layer. $L_{i, j}$ is the partial optical path of a photon within the $j^{th}$ tissue layer, and $L_f$ represents the partial optical pathlength through pulsating fetal artery. $p(L_1, L_2, \ldots, L_M )$ is now their joint PDF.

For simplification in the rest of the paper, we consider the sensed intensity is already normalized by the initial intensity, which is a common practice in diffuse optics. Rewriting equation \ref{eq:section2eq1} in a \emph{normalized} form gives:
\begin{equation} \label{eq:section2eq2}
    I = \int_{L_1} \int _{L_2} \ldots  \int_{L_f}, \ldots, \int_{L_M} \exp(-\sum_{j=1}^{M} \mu_{a,j} L_{j}) p(L_1, L_2, \ldots,  L_f, \ldots, L_M ) dL_1 dL_2 \ldots  dL_f, \ldots, dL_M
\end{equation}

This integral can be compressed into a single integral over all path lengths, denoted as $\fcyL$, to further tidy the equation:
\begin{equation} \label{eq:section2eq3}
    I = \int_{\fcyL} \exp(-\sum_{j=1}^{M} \mu_{a,j} L_{j}) p(\fcyL) d\fcyL
\end{equation}

There are many different ways of breaking down a human body into separate homogeneous tissue layers. The exact method used in our work is described in the following sections. For the purpose of this discussion, we assume one of the layers is composed of only fetal blood. The absorption coefficient($\mu_a$) of fetal blood is a function of both the concentration of fetal hemoglobin in the fetal blood layer, as well as fetal oxygen saturation.

Absorbance captures the combined effect of all the $\mu_a$'s within the measurement geometry. However, fetal blood oxygen saturation is specifically related to the fetal arterial $\mu_a$. One way to extract the fetal $\mu_a$ is to take advantage of pulsation. Arteries carry oxygenated blood to the extremities. As a result, the blood concentration within the fetal artery pulsates at the fetal heartrate (FHR). The impact of this fetal artery pulsation is visible in PPG waveform's frequency domain, provided we have a solid SNR. Many different techniques exist that can be used to isolate the peak and trough of the fetal pulsation\cite{bottrichSignalSeparationTransabdominal2018, kasapKUBAISensorFusion2023}. Conventionally, the extracted fetal peak and trough are used to calculate RoR and ultimately fSpO$_2$.

Let $I_1$ and $I_2$ be two points within the CW NIRS time series, such that they correspond to the peak and trough of a fetal pulsation. In other words, their difference corresponds to the amplitude of the fetal AC component. Let us temporarily assume that other tissue layers do not change between these points. Later, we discuss signal processing strategies to estimate $I_1$ and $I_2$ under the said assumption from \emph{in-vivo} data, where various physiological phenomenon, including maternal pulsation, respiration and Mayer waves are present. We define EPR as:
\begin{equation} \label{eq:section2eq4}
    \text{Exponential Pulsation Ratio(EPR)} = \frac{I_2}{I_1}
\end{equation}
Here, EPR is a non-logarithmic version of the classic Pulsation Ratio(PR) used in pulse oximetry. 

To further our derivations, we first define the absorption coefficients ($\mu_a$) for fetal blood at the peak and trough of pulsation are $\mu_{a,1}$ and $\mu_{a,2}$, respectively, while the $\mu_a$ values for other tissue layers remain constant. We also assume that there are no changes in the geometric or scattering coefficients between these two points, ensuring that the joint probability density function (PDF) of optical path lengths, $p(L_1, L_2, \ldots, L_M)$, is unchanged. The final assumption posits that the joint PDF is independent of the path length specific to the fetal blood layer ($p(L_f)$). Generally, this assumption might not hold in reflectance measurements as the path length distribution within any tissue layer can be influenced by other layers, through which the detected photons have traveled. However, given the small magnitude of fetal pulsation, this simplification does not practically introduce significant errors into the model.

Under these assumptions, we can simplify equation~\ref{eq:section2eq4} as:
\begin{equation} \label{eq:section2eq5}
    \begin{split}
    \text{EPR} & = \frac{\int_{\fcyL} \exp(-\sum_{j=1,j\neq f}^{M} \mu_{a,j} L_{j}) \exp(-\mu_{a,2}L_f) p(L_1, L_2, \ldots, L_f, \ldots, L_M) p(L_f) d\fcyL}{\int_{\fcyL} \exp(-\sum_{j=1,j\neq f}^{M} \mu_{a,j} L_{j}) \exp(-\mu_{a,1}L_f)p(L_1, L_2, \ldots,  L_f, \ldots, L_M) p(L_f)d\fcyL}  \\
    & = \frac{\int_{\fcyL} \exp(-\sum_{j=1,j\neq f}^{M}  L_{j})  p(L_1, L_2, \ldots, L_M) d\fcyL \int_{L_f}\exp(-\mu_{a,2}L_f) p(L_f)dL_f}{\int_{\fcyL} \exp(-\sum_{j=1,j\neq f}^{M} \mu_{a,j} L_{j}) p(L_1, L_2, \ldots, L_M) d\fcyL \int_{L_f}\exp(-\mu_{a,1}L_f) p(L_f)dL_f} \\
    & = \frac{\int_{L_f} \exp(-\mu_{a,2} L_f) p(L_f) dL_f}{\int_{L_f}\exp(-\mu_{a,1} L_f)p(L_f)dL_f} \\
    & = \frac{\EX_{L_f \sim p(L_f)}[\exp(-\mu_{a,2}L_f)]}{\EX_{L_f \sim p(L_f)}[\exp(-\mu_{a,1}L_f)]} \\
    \end{split}
\end{equation}

From equation \ref{eq:section2eq5}, it is evident that EPR depends solely on the fetal blood's $\mu_a$ and the optical pathlength distribution within the fetal layer, effectively isolating the signal from influences of other tissue layers. In other words, EPR provides us with the necessary signal isolation. This makes EPR an attractive feature for estimation of fetal blood oxygen saturation.


Importantly, the PDF term in equation~\ref{eq:section2eq5} is dependent on the patient anatomy, scattering properties, and the wavelength. This uniqueness to each patient renders the PDF non-trivial to determine, highlighting the limitation of conventional analytical methods. Consequently, machine learning-based approaches, which can adapt to complex data patterns, promise to offer a more effective solution for analyzing such patient-specific characteristics in fetal monitoring.

\subsubsection{Saturation-Geometry Ambiguity}
The saturation-geometry ambiguity stems from the fact that oxygen saturation estimation using a single detector is an under-determined problem. A single detector placed at a specific source-detector distance (SDD) captures two time-series signals from the two separate source wavelengths, each used to calculate two EPRs described by equation~\ref{eq:section2eq5}. EPR is a function of both $\mu_a$ and the PDF $p(L_f)$, which in turn is influenced by measurement geometry, scattering properties, and the wavelength.

In the single detector case, while we have two equations, we are asked to solve for two separate $\mu_a$ values as well as the PDF. Multiple solutions exist and often times they will also be physiologically valid. In case of TFO, two patients with two separate fetal oxygen saturation values, and different body geometries might yield identical EPR values.

Transitioning to a multi-detector setup introduces a significant advantage. Detectors placed at different SDDs interrogate slightly different tissue volumes. Although the corresponding PDFs differ, they are interconnected, adding much-needed depth and context to the problem. The relationship and mutual information among the detectors outputs enhances the determinacy of fSpO$_2$ inference, thereby providing a more robust framework. 

\subsection{Validation with Simulated Multi-Layer Tissue Model}\label{sec:val_mc}
\subsubsection{Monte Carlo Simulation Setup}
Monte Carlo simulation was utilized to obtain photon propagation information in numerous embodiment of the pregnancy tissue model discussed in Figure~\ref{fig:tissue_model}. To perform the simulations, we used a modified version of the GPU-accelerated MCXtreme software\cite{MonteCarloEXtreme2021}. MCXtreme simulates a large number of photons as they traverse through a 3D volume. The software supports placement of light sources and virtual spherical detectors within the simulated volume. The simulator works by tracking the path of virtual photons launched from the source until they are completely absorbed, scattered out of the model bounds, or hit a detector. The intensity sensed by each detector was then calculated by taking the summed intensity of all the photons that reached it.

We simulated two separate light sources of wavelengths 735nm and 850nm. Both sources are pencil beams of unit intensity, and are placed at the center of the setup, directly on top of the tissue model. Our choice of wavelength is different compared to conventional pulse oximetry. The wavelengths are specifically chosen, as they are previously shown to optimize low oxygen saturation estimation, applicable in fetal monitoring\cite{mannheimerWavelengthSelectionLowsaturation1997a}. The choice of wavelengths fulfilled two primary criteria simultaneously: (a) Similar fetal pathlength distributions across both wavelengths($p(L_f)$ equation~\ref{eq:section2eq5}); (b) High contrasting extinction coefficients($\epsilon$) for both oxy- and deoxy- hemoglobin.

Monte Carlo simulation noise is inversely proportional to the number of detected photons. As source-detector distance increases, photon count per detector is observed to drop exponentially. To combat this phenomenon, a number of detectors are placed in a set of concentric circles around the source. Due to symmetry in the tissue model, the received intensity at a particular SDD is obtained by averaging the sensed intensity of detectors on the circle with the same radius. Additionally, the number of detectors per circle is proportional to the circumference of the circle. To put it simply, outer circles contain more detectors. Combination of the two tricks results in reduction of noise in MC simulations, while maintaining a manageable computational overhead.

For further noise reduction, we use an interpolated version of EPR rather than the raw simulated EPR values. Despite the steps taken to reduce Monte Carlo simulation noise, there still exists a significant amount of noise in the tissue models with a deep fetal depth and at far detectors due to the low detected photon count. Since the simulation noise variance is inversely proportional to the detected photon count, reducing the noise requires an exponential increase in the number of photons simulated. This in turn requires a significant increase in computational resources. Furthermore, taking the fraction of two noisy values, as in the EPR calculation, only amplifies the noise. Thus we interpolate the EPR values to reduce computation time.

Our interpolation technique ensures smoothness in the EPR vs. SDD curve for each data point. We achieve this by taking the central difference for each EPR value in the EPR vs. SDD curve, taking a 2 point moving average over the central differences, and then regenerating the EPR vs. SDD curve using the smoothened central differences. All further analysis uses the interpolated EPR values. Interpolation directly improves machine learning model performance over using raw EPR values. Our interpolation method is not perfect. A more sophisticated, fetal depth-based interpolation method could be developed to squeeze out further performance improvements.

\subsubsection{Tissue Model}
\label{tissue_model}
We used a 4-layer flat, homogeneous, volumetric tissue model as shown in Figure \ref{fig:tissue_model}. Two aspects of the tissue model are varied to produce large quantities of simulation data. The first aspect is the hemodynamic parameters, which includes maternal hemoglobin concentration ($[Hb]_m$), maternal oxygen saturation ($s_m$), fetal hemoglobin concentration ($[Hb]_f$), and fetal oxygen saturation ($s_f$). The second aspect deals with geometric dimensions of the model. For this paper, we limit ourselves to varying the thickness of maternal abdominal wall layer ($d_m$), which creates tissue models with varying fetal depth.

\input{Sections/optical_prop_table.tex}

The optical properties and dimensions for each of the layers are given in Table~\ref{table:optical_prop} for both wavelengths of interest. These macro properties include the absorption coefficient($\mu_a$), scattering coefficient($\mu_s$), anisotropy factor($g$) and refractive index($n$). The absorption and scattering coefficients are wavelength-dependent. The anisotropy factor and refractive index are assumed to be constant across the two wavelengths. The model dimensions are modeled at term-pregnancy\cite{fongContextuallyawareFetalSensing2020}. Also, note that we added a thin amniotic fluid layer which does not significantly affect absorption but the scattering effect improves simulation results\cite{guntherEffectPresenceAmniotic2021a}.

To model optical changes due to pulsation, we assume that hemodynamic only affect $\mu_a$, leaving the rest of the optical properties, such as scattering coefficients, unchanged \cite{bosschaartLiteratureReviewNovel2014}. More specifically, maternal and fetal hemodynamics are assumed to only affect the $\mu_a$ of maternal abdominal wall and fetal tissue, respectively. 

The exact $\mu_a$ for the two pulsating layers are calculated as a summation of a static baseline non-blood $\mu_a$ and a dynamic blood $\mu_a$ term \cite{saidiTranscutaneousOpticalMeasurement1992, SkinOpticsSummary}. The baseline accounts for non-blood skin. However, we do not account for melanin. Moreover, We considered the contributions of arterial and venous blood separately. The $\mu_a$ for each type of blood was primarily dominated by total hemoglobin. Total hemoglobin could be further split into oxy- and deoxy-hemoglobin. Note that the percentage of oxyhemoglobin to total hemoglobin is the \textit{saturation}, ignoring the rest of the blood pigments. We also assumed that only 10\% of the overall tissue is blood in terms of volume, and that venous blood has 75\% less oxygen saturation compared to arterial blood. Whenever we refer to saturation in this paper, we always refer to arterial saturation, unless otherwise stated. The last assumption we made is that venous and arterial blood have equal blood volume fraction, i.e., in our case, 5\% each. Putting it all together, the overall tissue $\mu_a$ for Layer 1 and 4 is given by the following equation:

\begin{equation}
    \mu_{a,tissue} = 0.05 \times (\mu_{a,arterial blood} + \mu_{a,venous blood}) + 7.84 \times 10^7 \times wavelength ^{-3.255}
    \label{eq:tissue_mu_a}
\end{equation}

For both arterial and venous blood, the corresponding $\mu_a$ is given as $\text{Hb Concentration} \times (Saturation \times \epsilon_{HbO} + (1 - Saturation) \times \epsilon_{HHb})$, where $\epsilon_{HbO}$ and $\epsilon_{HHb}$ are the wavelength dependent extinction coefficients for oxy and deoxy hemoglobin respectively\cite{takataniTheoreticalAnalysisDiffuse1979}. 

Though MC simulations produced static intensity values, multiple simulation results could be stitched together to create a dynamic time series. This trick allows simulation of pulsation using multiple static points. For our purposes, we only required two temporal points, the fetal systole and diastole. In our simulations, these two points correspond to a 2.5\% change in Hb Concentration, while all other model parameters remain unchanged\cite{reussMultilayerModelingReflectance2005}. Note that we did not consider any geometric or scattering property changes during pulsation. Each pulsation point is simulated individually. The sensed intensity values are then used to calculate EPR.

\subsubsection{Simulation Acceleration}
We leveraged a mathematical trick to accelerate photon propagation simulations with CUDA. Recall that in our simulation setup, the hemodynamic parameters only affect $\mu_a$. MCXtreme can be instructed to store data for each photon individually, using which, optical pathlength within each tissue layer for each photon is stored in a large table. With this tabular data, equation~\ref{eq:section2eq1} enabled us to calculate the sensed intensity for any arbitrary set of $\mu_a$ value. This allowed us to change the $\mu_a$ for each tissue layer on the fly. We could then quickly gather data for a large number of $\mu_a$ combinations without re-running the MC part of the simulation. 

Note that the same approach cannot be applied to geometric parameters, such as the thickness of the maternal abdominal wall, which affect the optical pathlengths. Only a single MC simulation is required for each geometric parameter. One can then use the tabular data to calculate the sensed intensity for any combination of hemodynamic parameters. 

\subsection{Artificial Noise Injection}
\label{sec:method_artificial_noise}
We introduced artificial noise to our simulated photodetector data to evaluate the performance degradation of our neural network under more realistic, non-ideal conditions\cite{choeDiffuseOpticalTomography2005}. The pristine data generated by our tissue model represents a somewhat ideal scenario; however, various noise sources invariably affect real-world photodetector systems. By testing our trained network on datasets augmented with physiologically relevant noise, we can better estimate its performance in practical applications.

We focused on modeling the three main types of noise found in photodiode systems, especially those using trans-impedance amplifiers (TIAs). We model each noise as a noise current on the input side of the TIA.
\begin{enumerate}
    \item \textbf{Shot Noise:} Inherent quantization noise, $i_{shot} = \sqrt{2qBi_k}$; where $q$, $B$, and $i_k$ are electron charge, noise bandwidth, and generated photocurrent respectively.
    \item \textbf{Thermal Noise:} From TIA's gain resistor, $i_{thermal} = \sqrt{\frac{4kTB}{R}}$; where $k$, and $R$ are Boltzmann constant and gain resistor respectively. 
    \item \textbf{OpAmp Gain:} Which amplifies both noise current and signal current ($G$). In this hypothetical scenario, we choose a single set of gains over the entire dataset. The gains are chosen on a per-detector basis to normalize optical received power across all detectors.
\end{enumerate}
For simplicity, we ignore dark currents and other physiological noises originating from the patient body itself.

We consider two distinct scenarios. For the first scenario, we assume measurements are shot noise-dominated and ignore measurement noise. In the second scenario, we consider all three types of noise stated above. We compute the combined noise current in each scenario and promptly convert it into noisy light power($\sigma_{noise}$) via $\text{Generated Photocurrent} = \text{Responsivity} \times \text{Received Optical Power}$. We assume a responsivity of $0.60$ for all wavelengths, in line with our photodetector dataset. Ultimately, we created two additional noisy datasets, $I_{noisy}$, by injecting the computed noise power into the original dataset, $I_{original}$.
\begin{enumerate}
    \item \textbf{Scenario 1:} Shot-noise dominated, $I_{noisy, shot} = I_{original} + \mathcal{N}(0, \sigma_{shot})$
    \item \textbf{Scenario 2:} Shot \& Measurement noise, $I_{noisy, combined} = G \times (I_{original} + \mathcal{N}(0, \sqrt{\sigma_{shot}^2 + \sigma_{thermal}^2}))$
\end{enumerate}

For each combination, we apply a 5-iteration cross-validation and compute average MAE and standard deviation scores for both types of input data.

\subsubsection{fSpO$_2$ Estimation for Simulation Data}
\label{sec::sim_fetal_sat}
In each geometric constitution of the tissue model, four hemodynamic parameters were swept in Monte Carlo simulation so that a set of spatial intensity values representing certain subjects would be generated. These hemodynamic parameters include maternal hemoglobin concentration([Hb]$_m$), maternal oxygen saturation(S$_m$), fetal hemoglobin concentration([Hb]$_f$), and fetal oxygen saturation(S$_f$).The details of swept hemodynamic parameters in each tissue model parameter (TMP) are summarized in Figure~\ref{fig:tissue_model}. The number of samples from each depth is the combination of a set of sweep parameters, which results in 16,632 samples for each fetal depth and 515,592 samples in total from 31 fetal depth values. Each sample is a ten-features vector containing EPRs calculated from the spatial intensity at five detectors, and at two wavelengths. 

Specifically, each EPR relates to observation of a subject under certain physical state measured at a given SDD and wavelength. The geometry and hemodynamic parameters are normally unknown to the observer, which complicates the sensing objective, since finding an analytical function for estimation of fSpO$_2$ from multi-detector dual-wavelengths EPRs without knowledge of the geometry and hemodynamic parameters would be non-trivial. On the other hand, a data-centric approach through machine learning is compelling for information fusion across channels and wavelengths.

In this study, we adopt a Multi-Layer Perceptron (MLP) architecture for the task of fSpO$_2$ estimation. MLPs, originally introduced by Rosenblatt~\cite{Rosenblatt_1958}, are a widely used class of neural networks capable of modeling complex nonlinear relationships, and are typically trained using the backpropagation algorithm developed by Rumelhart et al.~\cite{Rumelhart_1986}. The MLP is well-suited for multi-sensor fusion in biomedical applications.

The MLP network used in this study takes an input vector $x\in \mathbb{R}^{1 \times 10}$, which is first passed through a hidden layer with $n$ neurons. The number of neurons in subsequent layers is reduced by a factor of two at each stage until reaching 8. A final linear layer with a single output node is used to generate the predicted continuous fSpO$_2$ value. Each hidden layer consists of a fully connected linear transformation, followed by a Rectified Linear Unit (ReLU) activation and a batch normalization layer to improve training stability.

The model is trained using the mean squared error (MSE) loss function to minimize the difference between predicted ($\hat{y}$) and reference fSpO$_2$ values ($y)$ and a corresponding MAE is also calculated as a direct reference of oxygen saturation estimation error:
\begin{equation} \label{sdr}
MSE := \frac{1}{n}\sum_{i=1}^{n}\left\| y_i - \hat{y_i}\right\|^2 \quad \text{and} \quad MAE := \frac{1}{n}\sum_{i=1}^{n}\left\| y_i - \hat{y_i}\right\|
\end{equation}
Optimization is performed using the Adam optimizer with a learning rate of $1\times10^{-3}$ and a weight decay of $1\times10^{-4}$. Training was conducted over a maximum of 300 epochs with a batch size of 32, and early stopping was applied with a patience of 25 epochs to prevent overfitting.

\subsection{Validation in Gold Standard Large Animal Model}\label{sec:val_sheep}
\subsubsection{Pregnant Ewe with In-Utero Hypoxic Fetal Lamb}
\label{sec::lamb_exp}

\begin{figure}[tbh!]
    \centering
    \begin{subfigure}[b]{0.15\textwidth}
        \centering
        \includegraphics[width=\textwidth]{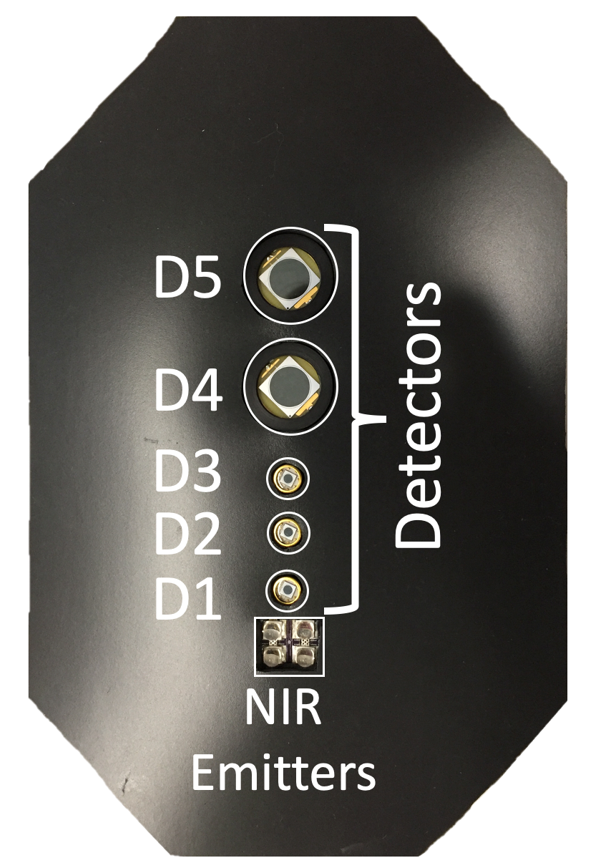}
        \caption{}
        \label{fig:device}
    \end{subfigure}
    \hfill
    \begin{subfigure}[b]{0.42\textwidth}
        \centering
        \includegraphics[width=\textwidth]{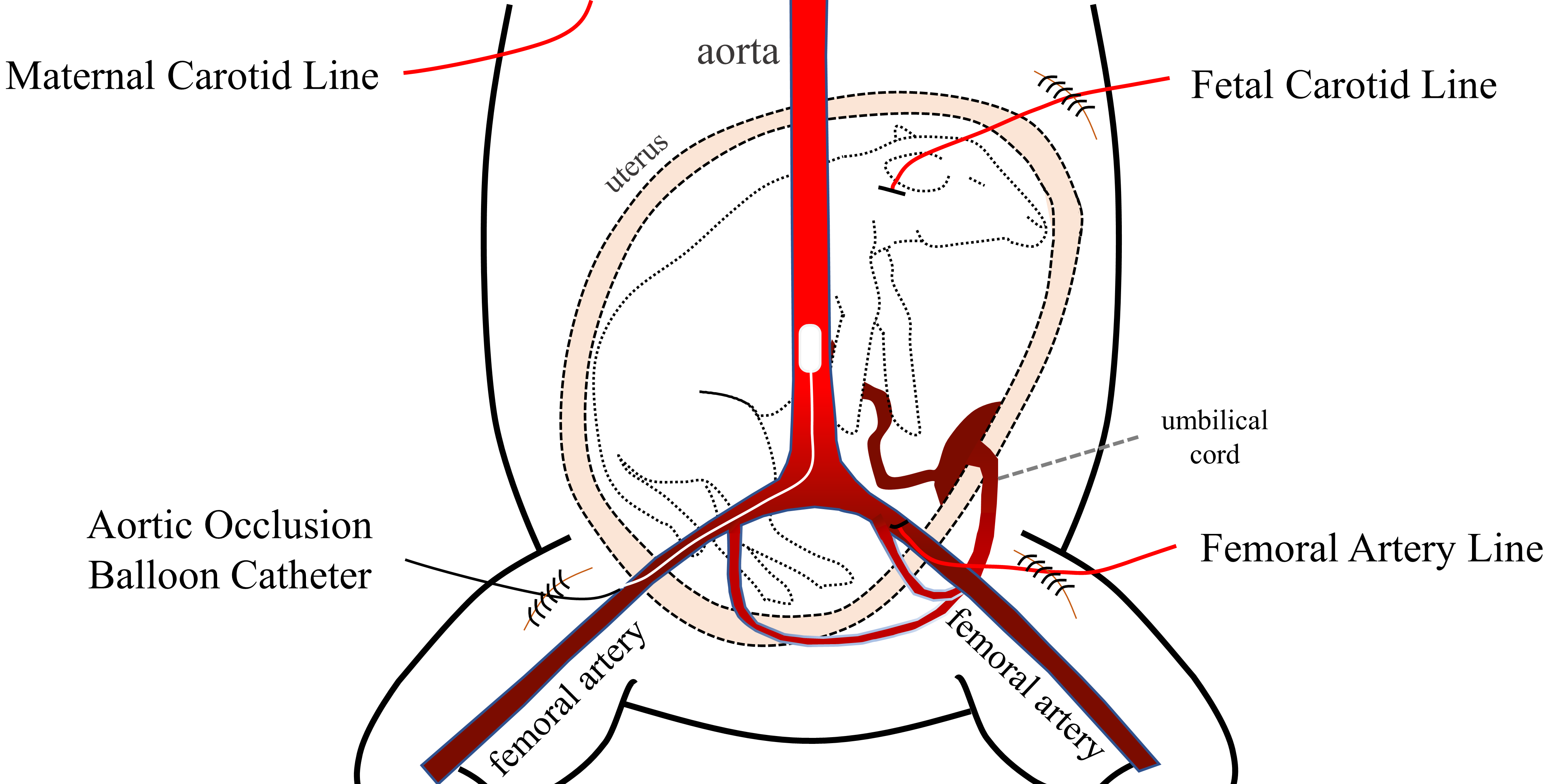}
        \caption{}
        \label{fig:sheep_exp}
    \end{subfigure}
    \hfill
    \begin{subfigure}[b]{0.42\textwidth}
        \centering
        \includegraphics[width=\textwidth]{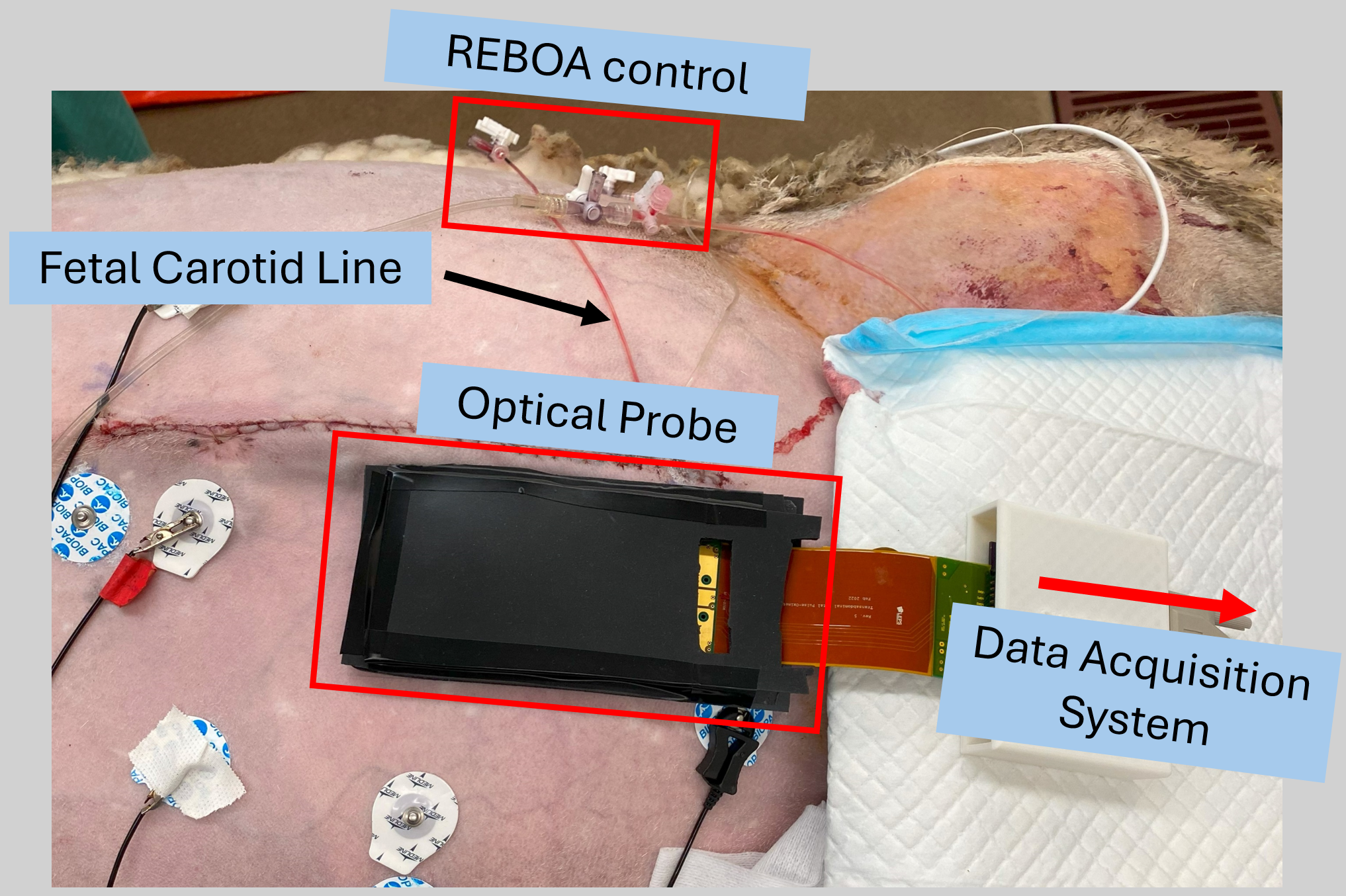}
        \caption{}
        \label{fig:surgery_pic}
    \end{subfigure}
    
    \vspace{1em}
    \begin{subfigure}[b]{0.58\textwidth}
        \centering
        \includegraphics[width=\textwidth]{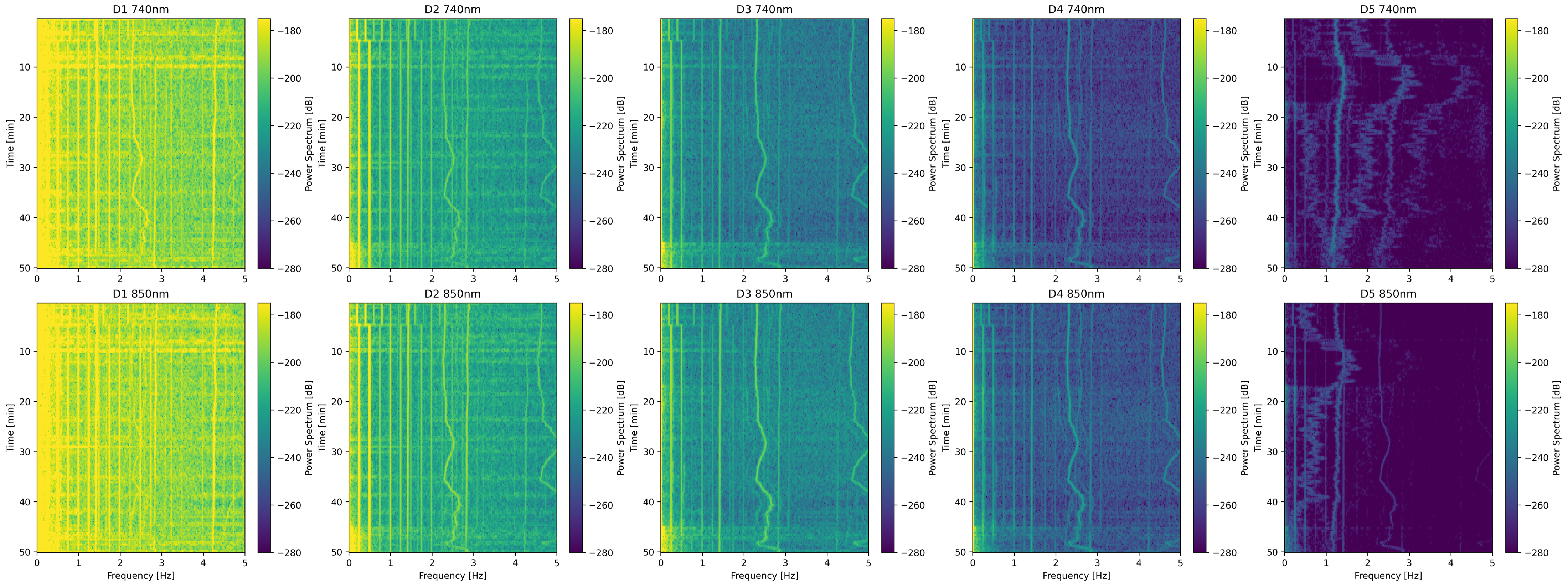}
        \caption{}
        \label{fig:sheep_spect}
    \end{subfigure}
    \hfill
    \begin{subfigure}[b]{0.38\textwidth}
        \centering
        \includegraphics[width=\textwidth]{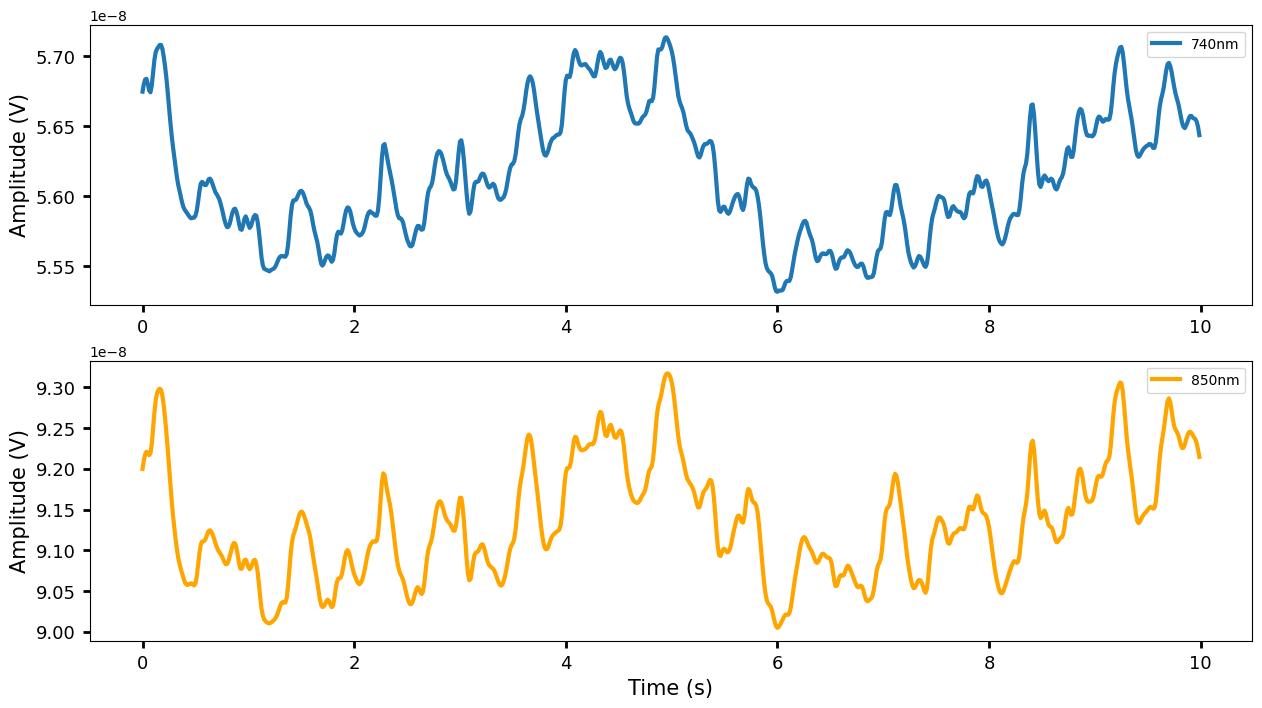}
        \caption{}
        \label{fig:sheep_ts}
    \end{subfigure}
\caption{\textbf{A controlled fetal hypoxia experiment on pregnant ewe with in-utero fetal lamb model was used to validate the TFO system and the usage of EPR for fSpO$_2$ estimation.} (a) TFO probe prototype features 5 optical detectors with SDD = \{1.5, 3, 4.5, 7, 10\}cm. (b) Sketch of controlled de-saturation experiment in pregnant ewe model\cite{Qian_CHASE}. (c) A photo from one experiment demonstrating the placement of TFO device on the ewe's abdomen. (d) Spectrograms representing the five-detectors dual-wavelength mixed PPG signals collected from a single hypoxic round. The upper row displays the spectrograms for the 740nm wavelength, while the lower row presents those for the 850nm wavelength. (e) A 10 seconds time-series from both wavelengths are shown.} 
\end{figure}

A pregnant ewe with in-utero hypoxic fetal lamb model was used to validate the TFO system. The study protocol was approved by UC Davis Institutional Animal Care and Use Committee (IACUC), numbered 22476, on September 17th, 2021. Five controlled fetal de-saturation experiments were conducted on time-mated term pregnant ewes with a mean gestational age of 136 days (approximately 12 days before nominal delivery time), and the resulting data were used in this study.

In each experiment, the pregnant ewe was anaesthetized with stable respiration controlled by external ventilator. An endovascular balloon catheter was placed through the femoral artery and advanced to the abdominal aorta to leverage Resuscitative Endovascular Balloon Occlusion of the Aorta (REBOA) technique. The aim is to induce fetal hypoxia via controlling the ewe’s aortic distal blood pressure through controllable inflation and partial occlusion of the ewe's aorta.

A hysterotomy was performed, and the fetal lamb was exteriorized. An arterial line was placed in the carotid artery of the lamb to monitor and record fetal hemodynamics. The line was also used to draw fetal blood samples. The fetus was returned to the uterus, and the fetal ear was sutured to the uterus to minimize the movement of the fetus in-utero. The lost amniotic fluid was replaced with warm saline, and the subsequently, the fetus and the ewe's abdominal layers were sutured back. The multi-detector optical probe was placed on the top of the ewe’s closed abdomen. The probe LEDs were positioned above the fetal head (Figure~\ref{fig:device}). A sketched of the controlled fetal de-saturation experiment is illustrated in Figure~\ref{fig:sheep_exp}.


The main goal of the hypoxic lamb experiment is to collect PPG data with TFO device while having access to reference fSaO$_2$ values. Fetal hypoxia was induced by inflating the aortic balloon catheter in steps, which in turn decreased ewe's distal Mean Arterial Pressure (dMAP). The reduced blood flow to the uterus induced hypoxia in the fetus. Each dMAP level was maintained for 10 minutes. Three blood samples were drawn from the fetus in each dMAP step at 2.5-, 5-, and 10-min, then the balloon catheter was inflated further for the next dMAP step. The fetal blood samples were analyzed via Arterial Blood Gas (ABG) analysis to obtain the reference fetal oxygen saturation (fSaO$_2$) data. Each hypoxic round stopped when two consecutive ABG readings were below 15\%. There were 45 minutes of recovery time planned between every two rounds in which, the balloon was completely deflated, and the fetus was given sufficient time for recovery. Two fetal blood samples were drawn at 30- and 45-min during the recovery, and the blood gasses were obtained. If the fSaO$_2$ was not observed to be above 15\% after 45-min of recovery time, then the experiment was deemed to have terminated, and euthanasia would be immediately performed. If the fetus recovered well, maximum of three hypoxic rounds would be performed, and both the ewe and fetal lamb would be euthanized after the third round.

\subsubsection{fSpO$_2$ Estimation in Lamb Experiment}
\label{sec::lamb_fSpO2}

The raw data captured by the TFO system consists of 5 channels of time-series waveforms sampled at 8,000 samples per second (SPS). Each channel contains the information from two different wavelengths of 740nm and 850nm. To separate the signals associated with each wavelength, the two LED light sources were frequency modulated by toggling the 740nm LED at 690Hz and the 850nm LED at 940Hz. Consequently, a frequency demodulation process was applied to the five detector channels, resulting in ten channels of single-wavelength PPG data. To reduce the data size, improve SNR and computational efficiency, the demodulated signal was down-sampled to 80Hz. From five hypoxic lamb experiments, each containing two or three rounds, eight rounds were selected after inspecting the SNR of the fetal pulsation in the acquired PPG data.

The collected PPG signal exhibited a wide range of power, mixed with various signal components. The main components in the acquired signal were maternal respiration, maternal pulsation, and fetal pulsation. The empirical ranges for pregnant ewe models are 1.5-3.5Hz (90-210bpm) for Fetal Heart Rate (FHR), 1.1-2Hz (72-120bpm) for Maternal Heart Rate (MHR), and 0.2-0.33Hz (12-20bpm) for Maternal Respiratory Rate (MRR). Thus, the collected PPG is a mixture of quasi-periodic signals with frequency-varying maternal and fetal components due to dynamic pulsation rates. 

To capture the EPRs from multi-detector PPG signals collected during experiment, lower signal enveloping and lock-in detection methods were employed to separate the DC and fetal AC components, respectively. The trough intensity (DC) represents a stable wave originating from the static light attenuation by non-pulsatile chromophores in maternal and fetal tissue, which is captured by the lower signal envelope in our analysis (systole intensities, $I_1$). A signal envelope outlines and interpolates the signal's extremes, effectively discarding higher frequency components.

To extract the peak intensity, we used lock-in detection, a technique that isolates the amplitude of a signal at a known frequency buried in noise. Continuous FHR data was recorded during each experiment via monitoring of blood pressure dynamics through the fetal carotid arterial line via a pressure transducer. We approximate the peak-to-trough intensity change due to fetal pulsation using the peak-to-peak amplitude of a sine wave with the same frequency as the FHR, determined from the PPG signal spectrum at the FHR frequency (Figure \ref{fig:overview}). This approximation, commonly used in conventional single-body pulse oximetry, has been shown to provide reasonably accurate results \cite{Rusch_1996}.

The sum of the DC and two times the AC components yielded the diastole intensities ($I_2$). Subsequently, the EPR for each specific detector and wavelength pair was approximated as EPR = (2*AC$_{fetal}$ + DC) / DC (Figure~\ref{fig:overview}). To reduce noise, a moving average filter with a 1.5-minute window length was applied to the EPRs series.

Compared to conventional pulse oximetry, the presence of strong, non-stationary and complex \emph{physiological clutter} originating in the superficial tissue layers, significantly challenge the sensing objective. The complexity arises in part from the exponential impact of depth on strength of the fetal layer signal, which results in highly diminished signal of interest relative to the various noise sources contributing to the measured data. Detector fusion is a promising approach to counter this problem, as the information collected by each detector is highly correlated. Analogous to fSpO$_2$ estimation for simulation data, described in section~\ref{sec::sim_fetal_sat}, a MLP architecture is employed to fuse information across detectors in animal experiments.

The validation scheme of fSpO$_2$ estimation involves a temporal split of data for training and validation data. A random split of EPR samples would lead to information leakage between training and validation sets, as adjacent EPR samples relate to overlapping PPG data. Therefore, each hypoxic round was partitioned into five contiguous folds: one fold for validation and four folds for training. Additionally, each hypoxic round exhibits a similar trend of decreasing fSaO$_2$ due to the progressive inflation of the endovascular balloon in the study protocol. To reduce the associated bias in label distribution, for each iteration a different fold is used for validation, ensuring that both training and validation samples contain a representative set of examples at varying fSaO$_2$ levels.

Similar to the fSpO$_2$ estimation using the simulation dataset, MSE was used as the loss function, with MAE included as an additional metric to provide a more intuitive interpretation of training and validation errors. The optimizer and other training strategies were kept consistent with those described in section~\ref{sec::sim_fetal_sat}. Since the duration of each hypoxic round is different, the number of samples from each sensing geometry and label has a non-uniform distribution. To counter the issue, a weight inversely proportional to the number of samples was applied to samples from different rounds. This technique allows the network to value each sensing geometry equally in the training process, and prevents undue bias towards longer rounds from which, more samples are available. The initial weights of the neural network were randomized using a normal distribution, while the initial bias values were set to zero. Similar to the training and validation for the simulation dataset, the training process was scheduled for 300 epochs, incorporating an early stopping strategy to halt training if the validation error did not improve for 25 epochs, with the best model weights restored for inference.


%% file: Sections/optical_prop_table.tex
\begin{table}[]
    \centering
    \begin{tabular}{c|cc|cc|c|c|c}
\hline
\multirow{2}{*}{\textbf{Tissue Layer}} & \multicolumn{2}{c|}{\textbf{735nm}}                                                                                                                                  & \multicolumn{2}{c|}{\textbf{850nm}}                                                                                                                                  & \multirow{2}{*}{\textbf{\begin{tabular}[c]{@{}c@{}}Anisotropy\\ (g)\end{tabular}}} & \multirow{2}{*}{\textbf{\begin{tabular}[c]{@{}c@{}}Refractive\\ Index\\ (n)\end{tabular}}} & \multirow{2}{*}{\textbf{\begin{tabular}[c]{@{}c@{}}Thickness\\ (cm)\end{tabular}}} \\ \cline{2-5}
                                       & \multicolumn{1}{c|}{\textbf{\begin{tabular}[c]{@{}c@{}}$\mu_a$ \\ $mm^{-1}$\end{tabular}}}  & \textbf{\begin{tabular}[c]{@{}c@{}}$\mu_s$\\ $mm^{-1}$\end{tabular}}   & \multicolumn{1}{c|}{\textbf{\begin{tabular}[c]{@{}c@{}}$\mu_s$ \\ $mm^{-1}$\end{tabular}}}  & \textbf{\begin{tabular}[c]{@{}c@{}}$\mu_s$ \\ $mm^{-1}$\end{tabular}}  &                                                                                    &                                                                                            &                                                                                    \\ \hline
Maternal Abdomen                       & \multicolumn{1}{c|}{Eq.~\ref{eq:tissue_mu_a}}                       & 11.816\cite{simpsonNearinfraredOpticalProperties1998} & \multicolumn{1}{c|}{Eq.~\ref{eq:tissue_mu_a}}                       & 11.169\cite{simpsonNearinfraredOpticalProperties1998} & 0.9                                                                                & 1.4                                                                                        & 0.4 - 3.4                                                                          \\ \hline
Uterus Wall                            & \multicolumn{1}{c|}{0.0158\cite{simpsonNearinfraredOpticalProperties1998}} & 10.575\cite{simpsonNearinfraredOpticalProperties1998} & \multicolumn{1}{c|}{0.0991\cite{simpsonNearinfraredOpticalProperties1998}} & 8.125\cite{simpsonNearinfraredOpticalProperties1998}  & 0.9                                                                                & 1.4                                                                                        & 0.5                                                                                \\ \hline
Amniotic Fluid                         & \multicolumn{1}{c|}{0.0125\cite{modenaAmnioticFluidDynamics2004}}          & 0.1\cite{modenaAmnioticFluidDynamics2004}             & \multicolumn{1}{c|}{0.0042\cite{modenaAmnioticFluidDynamics2004}}          & 0.1\cite{modenaAmnioticFluidDynamics2004}             & 0.9                                                                                & 1.33                                                                                       & 0.1                                                                                \\ \hline
Fetal Tissue                           & \multicolumn{1}{c|}{Eq.~\ref{eq:tissue_mu_a}}                       & 12.5\cite{ijichiDevelopmentalChangesOptical2005}      & \multicolumn{1}{c|}{Eq.~\ref{eq:tissue_mu_a}}                       & 9.916\cite{ijichiDevelopmentalChangesOptical2005}     & 0.9                                                                                & 1.4                                                                                        & Semi-infinite                                                                      \\ \hline
\end{tabular}%
    \caption{\textbf{Optical properties used in multi-layer tissue model simulation.}}
    \label{table:optical_prop}
\end{table}

%% file: Sections/5_Data_Availability.tex
\label{sec:data_avail}

The data used in this study were obtained at University of California, Davis. Data is available upon request.

%% file: Sections/6_Code_Availability.tex
\label{sec:code_avail}

The code for training and evaluating the models used in this work is available at https://github.com/Garyqwt/TFO$\textunderscore$EPR.

%% file: Sections/7_Acknowledgement.tex
\label{sec:acknowledgement}
This work was supported by the National Science Foundation (NSF) under Grants IIS-1838939, CCF-1934568; the National Institutes of Health (NIH) under Grant 5R21HD097467; the National Center for Interventional Biophotonic Technologies (NCIBT) through NIH Grant 1P41EB032840; and the University of California-Noyce initiative.

%% file: Sections/8_Author_Information.tex
\label{sec:author_info}
\subsection{Authors and Affiliations}
\noindent

\textbf{Department of Electrical and Computer Engineering, University of California, Davis, Davis, CA, 95616, USA}

Weitai Qian, Rishad R. Joarder, Randall Fowler, Begum Kasap, Mahya Saffarpour, Kourosh Vali, Tailai Lihe, and Soheil Ghiasi

\textbf{Department of Surgery, UC Davis Health, Sacramento, CA, 95817, USA}

Aijun Wang and Diana Farmer

\textbf{Department of Biomedical Engineering, University of California, Davis, Davis, CA, 95616, USA}

Aijun Wang

\subsection{Contributions}
W.Q., R.R.J., and S.G. made substantial contributions to the study conception and design, manuscript preparation and revision. W.Q. and R.R.J. performed data analysis and visualization. W.Q., R.R.J., R.F., B.K., M.S., K.V., and T.L. contributed to the data collection and pre-processing. A.W., and D.F. provided clinical supervision in the study. S.G. provided academic supervision on study design, reviewing and editing manuscript.

%% file: Sections/9_Ethics_Declarations.tex
\label{sec:ethic_dec}
\subsection{Competing interests}
Soheil Ghiasi is a co-founder of Storx Technologies, an early stage spinoff, which aims to commercialize the TFO technology. Other authors declare no competing interests.